\documentclass{aa}
\usepackage{txfonts}
\usepackage{graphicx}
\usepackage{amsmath}

\newcommand{\msol}{\hbox{\kern 0.20em $M_\odot$}}

\newcommand{\lsol}{\hbox{\kern 0.20em $L_\odot$}}
\newcommand{\g}{\hbox{\kern 0.20em g}}
\newcommand{\gmu}{\hbox{\kern 0.20em g$^{-1}$}}
\newcommand{\kg}{\hbox{\kern 0.20em kg}}
\newcommand{\pc}{\hbox{\kern 0.20em pc}}
\newcommand{\mum}{\hbox{\kern 0.20em $\mu$m}}
\newcommand{\mumd}{\hbox{\kern 0.20em $\mu$m$^{-2}$}}
\newcommand{\cm}{\hbox{\kern 0.20em cm}}
\newcommand{\m}{\hbox{\kern 0.20em m}}
\newcommand{\km}{\hbox{\kern 0.20em km}}
\newcommand{\nm}{\hbox{\kern 0.20em nm}}
\newcommand{\s}{\hbox{\kern 0.20em s}}
\newcommand{\h}{\hbox{\kern 0.20em h}}
\newcommand{\smu}{\hbox{\kern 0.20em s$^{-1}$}}
\newcommand{\smd}{\hbox{\kern 0.20em s$^{-2}$}}
\newcommand{\an}{\hbox{\kern 0.20em an}}
\newcommand{\anmu}{\hbox{\kern 0.20em an$^{-1}$}}
\newcommand{\yr}{\hbox{\kern 0.20em yr}}
\newcommand{\yrmu}{\hbox{\kern 0.20em yr$^{-1}$}}
\newcommand{\Myr}{\hbox{\kern 0.20em Myr}}
\newcommand{\Mymu}{\hbox{\kern 0.20em Myr$^{-1}$}}
\newcommand{\K}{\hbox{\kern 0.20em K}}

\newcommand{\pcmu}{\hbox{\kern 0.20em pc$^{-1}$}}
\newcommand{\pcmd}{\hbox{\kern 0.20em pc$^{-2}$}}
\newcommand{\pcmt}{\hbox{\kern 0.20em pc$^{-3}$}}
\newcommand{\kms}{\hbox{\kern 0.20em km\kern 0.20em s$^{-1}$}}
\newcommand{\kmpd}{\hbox{\kern 0.20em km$^{2}$}}
\newcommand{\kpc}{\hbox{\kern 0.20em kpc}}
\newcommand{\cms}{\hbox{\kern 0.20em cm\kern 0.20em s$^{-1}$}}
\newcommand{\erg}{\hbox{\kern 0.20em erg}}
\newcommand{\ergs}{\hbox{\kern 0.20em erg}}
\newcommand{\cmpd}{\hbox{\kern 0.20em cm$^2$}}
\newcommand{\cmmd}{\hbox{\kern 0.20em cm$^{-2}$}}
\newcommand{\cmms}{\hbox{\kern 0.20em cm$^{-6}$}}
\newcommand{\cmpt}{\hbox{\kern 0.20em cm$^3$}}
\newcommand{\cmmt}{\hbox{\kern 0.20em cm$^{-3}$}}
\newcommand{\mpd}{\hbox{\kern 0.20em m$^2$}}
\newcommand{\mmd}{\hbox{\kern 0.20em m$^{-2}$}}
\newcommand{\mpt}{\hbox{\kern 0.20em m$^3$}}
\newcommand{\mmt}{\hbox{\kern 0.20em m$^{-3}$}}
\newcommand{\mujy}{\hbox{\kern 0.20em $\mu$Jy}}
\newcommand{\mjy}{\hbox{\kern 0.20em mJy}}
\newcommand{\Mj}{\hbox{\kern 0.20em MJy}}
\newcommand{\jy}{\hbox{\kern 0.20em Jy}}
\newcommand{\ghz}{\hbox{\kern 0.20em GHz}}
\newcommand{\G}{\hbox{\kern 0.20em G}}
\newcommand{\muG}{\hbox{\kern 0.20em $\mu$G}}

\newcommand{\htwo}{\hbox{H${}_2$}}

\begin{document}

   \title{H$_2$ mass--velocity relationship from 3D numerical \\ 
         simulations of jet-driven molecular outflows}


   \author{A.H. Cerqueira\inst{1},
          B. Lefloch\inst{2}
                    \and
          A. Esquivel\inst{3}
          \and 
          P. R. Rivera-Ortiz\inst{2}
          \and 
          C. Codella\inst{4,2}
         \and
          C. Ceccarelli\inst{2}
          \and
          L. Podio\inst{4}
          }

   \institute{LATO/DCET, Universidade Estadual de Santa Cruz,
              Rod. Jorge Amado, km 16, Ilh\'eus, BA, CEP 45662-900, Brazil \\
              \email{hoth@uesc.br} \\
         \and
             CNRS, IPAG, F-38000 Grenoble, France \\
              Univ. Grenoble Alpes, IPAG, BP 53, F-38041 Grenoble, France \\
              \email{bertrand.lefloch@obs.ujf-grenoble.fr} \\
         \and
             Instituto de Ciencias Nucleares, Universidad Nacional Aut\'onoma de M\'exico, \\
        Apartado Postal 70-543, 04510 Ciudad de M\'exico, M\'exico \\
         \and
         INAF, Osservatorio Astrofisico di Arcetri, Largo E. Fermi 5, 50125 Firenze, Italy
           }

   \date{Received: \_\_\_\_\_\_ ; accepted: \_\_\_\_\_\_}

\titlerunning{Jet-cavity kinematical relationship}
\authorrunning{Cerqueira et al.}
  \abstract
{Previous numerical studies have shown that in protostellar outflows,
the outflowing gas mass per unit velocity, or mass--velocity
distribution $m(v)$, can be well described by  a broken power law
$\propto v^{- \gamma}$. On the other hand, recent observations of
a sample of outflows  at various stages of evolution show that the
CO intensity--velocity distribution, closely related to $m(v)$,   follows an exponential law $\propto \exp(-v/v_0)$.}
{In the present work, we revisit the physical origin of the mass--velocity
relationship $m(v)$ in jet-driven protostellar outflows. We investigate
the respective contributions of the different regions of the outflow,
from the swept-up ambient gas to the jet.}
{We performed 3D numerical simulations of a protostellar jet
propagating into a molecular cloud using the hydrodynamical code
Yguaz\'u-a.  The code takes into account the most abundant atomic and ionic
species and was modified to include the H$_2$ gas heating and
cooling.}
{We find that by  excluding the jet contribution, $m(v)$ is
satisfyingly fitted with a single exponential  law, with $v_0$ well
in the range of observational values. The jet contribution results
in additional components in the mass--velocity relationship.
This empirical mass--velocity relationship is found to be valid
locally in the outflow. The exponent  $v_0$ is almost constant in
time  and for a given level of mixing between the ambient medium
and the jet material.  In general, $v_0$ displays only a weak spatial
dependence.  A simple modeling of the L1157 outflow successfully
reproduces the various components of the observed CO intensity--velocity
relationship. Our simulations indicate that these components trace
the outflow cavity of swept-up gas and the material entrained along
the jet, respectively.}
{The CO  intensity--velocity exponential law is  naturally explained
by the jet-driven outflow model. The entrained material plays an
important role in shaping the mass--velocity profile.}
\maketitle
%

\section{Introduction}\label{int}

Outflows from young stellar objects (YSOs) can exhibit a great
variety of morphological and physical characteristics.  In the
youngest ($10^4 \yr$) and deeply embedded Class 0 protostars
\citep{andre}, outflows are easily traced using the CO molecule and
their presence is ubiquitous in star forming regions, indicating
that they are a common manifestation of both low- and high-mass
star formation processes \citep{wu,lee20}.  On the other hand,
protostellar jets were first associated with more evolved
Class II objects, that is, optically revealed pre-main sequence
objects that are still accreting (or classical T-Tauri stars). These
jets are observed mainly through forbidden atomic emission lines,
like [S II] and [N II], as well in H$\alpha$ \citep{rei01}.  In
between these two limiting cases, Class I protostars, with a typical
age of $10^5 ,$ may show evidence for both molecular outflows and
protostellar jets at the same time \citep[e.g., L1448 IRS 2 and IRS
3, see][]{bally97}.  Sometimes a fast and collimated molecular jet
is also observed, as in Cep E-mm \citep{lefloch15} or HH 212, which
are associated with a Class 0 source \citep[][]{zinnecker,lee17}.

Nevertheless, the origin of the molecular outflows associated with YSOs is still debated \citep[see][for a recent review]{lee20}.  
They are believed to be the by-product
of an interaction between a more collimated jet and/or wind, produced
in or by the star--disk interaction, and its surrounding medium
\citep{bally}. As the wind and/or jet bow shock propagate into the
ambient medium, the gas of the excavated cavity walls advances and
excites a profusion of molecular emission lines. For low mass YSOs in particular, this can take place via one of three main
mechanisms \citep[see][for a comprehensive review]{arce}: (i) 
In {\it wind-driven shell models},  a wide-angle wind is supposed
to accelerate the ambient medium gas. In this class of models, both
the wind and the surrounding medium are assumed to be
stratified in density. (ii) In {\it turbulent jet models,} 
a jet subject to dynamical Kelvin-Helmholtz instability can entrain
gas through a growing turbulent layer, giving rise to an outflow.
This mechanism can also operate at the leading working surface.
(iii) In {\it Jet bow-shock models,}  a collimated jet produces a
leading bow shock that accelerates the ambient medium gas. Also,
an intermittent jet may develop a set of internal working surfaces
that can help in the process \citep{raga1993}.

As emphasized in \cite{arce}, it is possible that more than one
mechanism is operating to produce a given molecular outflow, or
alternatively a given mechanism can dominate at different epochs
in the evolution of a given source.  In any case, a parameter that
has been historically used to identify useful models is the
slope of a power law that relates the mass of the outflowing molecular
gas with its velocity, or $m(v) \propto v^{-\gamma}$.  Rigorously
speaking, the mass--velocity relationship, sometimes called the mass
spectrum, is obtained by considering the mass in a given radial
velocity bin, meaning that the observed relationship is actually $\delta
m(v)/\delta v \propto v^{-\gamma}$ \citep[see][for a discussion]{arce2001}.
However, for the sake of simplicity, some authors refer to the
mass--velocity relationship as $m(v) \propto v^{-\gamma}$
\citep{DR99,downes}. The mass--velocity relationship gives us
the mass of the outflow at a given radial velocity. We note
that what is actually observed is the intensity of a given emission
line, typically the $^{12}$CO (1-0) line profile, and that such
an intensity correlates with the velocity as described above
\citep{DR99,arce2001}. The intensity is then converted to mass
(corrected or not by the opacity) to finally obtain the mass--velocity
relationship.  Molecular outflows seem to display a mass--velocity
relationship that can be described by a broken power law (Bachiller
\& Tafalla 1999; Ridge \& Moore 2001; Arce \& Good\-man 2001), with
shallower slopes ($\gamma < 2$) at low velocities ($v < 10 \kms$) and
steeper slopes ($\gamma > 3$) at intermediate-to-high velocities
($v > 10 \kms$).  In this way, no matter the mechanism used to model a
molecular outflow, the model should account for the observed slopes.

In the present paper, we focus on the molecular mass--velocity
relationship for molecular outflows produced by a collimated and
supersonic jet using three-dimensional numerical simulations. As
the jet interacts with the ambient medium, jet-entrained gas and ambient gas 
 swept up by the jet-driven bow-shock can in principle be disentangled.
The appeal of such a scenario is two-fold: (i) there is
increasing evidence that both phenomena may coexist in Class 0 and
Class I sources, as mentioned in the previous paragraph, and
(ii) molecular outflows produced by either jet entrainment or a jet-driven bow-shock can effectively end up in a power-law mass--velocity
relationship (Chernin \& Mason 1993; Zhang \& Zheng 1997; Stahler
\& Palla 2004). In the following section, we briefly compile some previously
important results obtained through numerical simulations of jet-driven
molecular outflows, and discuss some recent observational findings
that ultimately motivated the present work.

\section{Previous numerical results and observed 
intensity--velocity relationships}\label{newsec2}

Numerical simulations of molecular jets have been used extensively
 in the literature as an efficient tool to investigate
the kinematical properties of jet-driven molecular outflows
\citep[see][]{DR99, downes,rs03,rs04,sr05,sr07}. The mass--velocity
and intensity--velocity relationships $I_{\rm CO}(v)$ observed
in low-J ($\leq 8$) CO lines in molecular outflows have been studied
by various authors as a possible test for discriminating between
entrainment mechanisms \cite[see][hereafter, DC03]{downes}.

Previous works have described the CO intensity--velocity distribution
$I_{\rm CO}(v)$ in outflows as a broken power law, $I_{\rm CO}(v)\propto
v^{-\gamma}$ with $\gamma \simeq 1.8$ up to line-of-sight velocities
$v_{\rm break}\approx $10 -- 30 $\kms$ and a steeper slope $\gamma
= 3$ -- 7 at higher velocities \cite[see][for a review]{frank14}.
The CO intensity--velocity distribution behavior was successfully
reproduced by HD simulation of jet-driven flows, and was found to be the result of CO
dissociation above shock speeds of $\sim 20\kms$ and of the temperature
dependence of the line emissivity (see DC03).

\cite{DR99} introduced the H$_2$ molecule in their calculations
and found that the \htwo\ mass--velocity distribution $m(v) \propto
v^{-\gamma}$ follows a similar relationship to the intensity--velocity
relationship observed in the millimeter rotational lines of CO.
\cite{downes} showed that the swept-up molecular gas follows a
mass--velocity relationship ($m_{{\rm H}_2}(v)$)  similar  to the
intensity--velocity relationship $I_{{\rm H}_2}(v)$ in the low
velocity range ($v \lesssim 30$ $\kms$). In contrast, in the high
velocity range these latter authors found that $I_{{\rm H}_2}(v)$ is shallower than
$m_{{\rm H}_2}(v)$, while $I_{{\rm CO}}(v)$ is steeper than $I_{{\rm
H}_2}(v)$ but comparable to $m_{{\rm H}_2}(v)$.

\cite{rs03} focused on time-dependent jets, that is, jets whose
density varies as a function of time with respect to the ambient
medium. These latter authors found that the mass--velocity distribution is systematically
shallower than the CO intensity--velocity distribution. They also
found that the indices of the distributions are essentially unchanged
when considering atomic or molecular jet material.

\cite{keegan}  studied the temporal evolution of the power index
$\gamma,$ and their results are consistent with those of \cite{downes}.
Interestingly, Keegan and Downes  found that $\gamma$  should increase slowly
in time, attaining a limiting value after $t \approx 1500$ years.

We note that simulations and observational work have focused on the
global  properties (mass, etc.) and not on local properties inside
the outflows.  Also, the underlying bow-shock model predicts a power law
at low velocities on long computational timescales much longer
than the dynamical timescales derived  from observations, which are
usually a few thousand years \cite[see also][]{rs03,rs04,sr05,sr07}.

The discovery that the CO line profiles observed towards the
protostellar outflow L1157 could be very well fitted by an exponential
law $I_{\rm CO}(v)\propto \exp(-v/v_0)$, with $v_0 \sim 2-12 \kms$,
came as a surprise \citep{Lefloch2012}.  Further observational
studies confirmed that these spectral signatures, with similar
values  of $v_0$, were detected in a plethora of molecular gas
tracers, like CS \citep{gr2015}, HNCO and NH$_2$CHO \citep{mendoza2014},
HC$_3$N \citep{mendoza2018}, and HCO$^{+}$ \citep{podio2014}.

The same analysis applied to the outflow sample of \cite{bachiller99}
observed in the CO $J$=2--1 line (L1448, Orion A, NGC2071, L1551
and Mon R2) yielded a similar conclusion. The sample contains sources
at various stages of evolution from early Class 0 (e.g. L1448) to
late Class I  (e.g. L1551). \cite{Lefloch2012} showed that all the
sources could be fitted by a single exponential $I_{\rm CO}(v)\propto
\exp(-v/v_0)$, with values of $v_0$ between 2 and $12\kms$, well
in the range of those determined in L1157-B1.  Also, Lefloch et al. observed
the trend that the more evolved, Class I outflows of the sample
(Mon R2, L1551) display a shallower intensity--velocity distribution.
Therefore, an exponential relation $I_{\rm CO}(v) \propto \exp(-v/v_0)$
is found to be a good approximation of the observed intensity relation
not only in L1157-B1 but in several molecular outflows in general,
with a reduced number of free parameters compared to a broken power
law.  Mapping of the CO $J=$ 3--2 emission over  the L1157 outflow
by \cite{Lefloch2012} showed that the exponential spectral signature
is detected along the outflow cavity walls in the southern outflow
lobe, implying that it is not only a global, but also a local
signature of the outflowing gas.

In summary, molecular outflowing gas shows an intensity--velocity
relationship well described by an exponential power law, both on a
global and local scale. The question arises as to how this general property
of molecular outflows is related with the outflow mechanism itself.
In order to address this question, we carried out three-dimensional (3D)
hydrodynamics (HD) simulations of the evolution of ouflowing gas
in a dense medium representing the parental molecular cloud/surrounding
protostar envelope.  Specifically, we present a new methodology
based on distinguishing the mixing level between the ambient medium
and the jet gas, which helped us to disentangle the distinct components
that arise in the H$_2$ mass--velocity relationship.  The article
is organized as follows.  In \S \ref{setup} and \S \ref{simu} we
provide details of the numerical setup and initial parameters for
the simulations. In \S \ref{comparison} we briefly compare our
results with some previous numerical studies of molecular jets.  In
\S \ref{results} and \ref{discussion} we present the results of our
numerical simulations and we discuss their implications for
observations of molecular outflows using L1157 as a reference.  In
\S \ref{conc} we present our conclusions.

\section{Numerical simulations}\label{numeric}

\subsection{Numerical setup}\label{setup}

The simulations presented here were performed using the Yguaz\'u-a
code \citep{2000RMxAA..36...67R,raga2002, 2006A&A...448..231C}.  In
its original version, the code was designed to solve hydrodynamic
problems with a chemical network for the following atomic and ionic
species: HI, HII, HeI, HeII, HeIII, CII, CIII, CIV, NI, NII, NIII,
OI, OII, OIII, OIV, SII and SIII.  For the present  work, we
introduced the H$_2$ molecule as a new species and added three
dissociation reactions for molecular hydrogen:

\begin{equation}\label{dissoc1}
\rm H+H_2 \rightarrow 3H\, {\rm ,}
\end{equation}

\begin{equation}\label{dissoc2}
\rm H_2+H_2 \rightarrow 2H + H_2\, {\rm ,}
\end{equation}

\begin{equation}\label{dissoc3}
\rm e + H_2 \rightarrow e + 2H \, {\rm .}
\end{equation}

\begin{figure}
\centering
\includegraphics[width=\columnwidth]{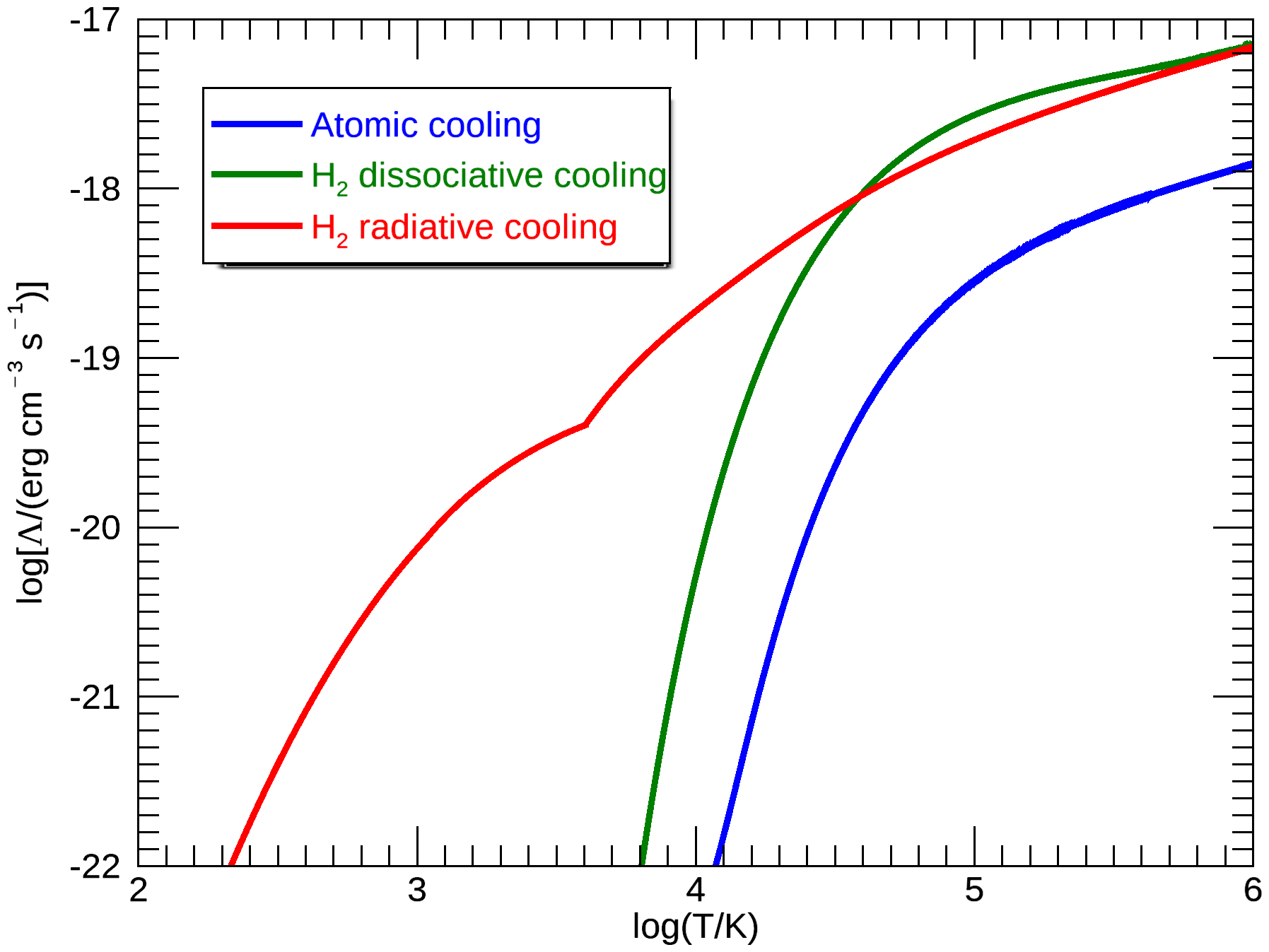}
\caption{Different contributions for the cooling: atomic emission
lines (blue line), H$_2$ dissociative cooling (green line), and H$_2$
radiative cooling (red line).  The cooling functions were calculated
using the starting values (i.e., at $t=0$) for the numerical
densities, or $n_{{\rm HI}}$ = 8.31 cm$^{-3}$, $n_{{\rm HII}} =
0.08$ cm$^{-3}$, $n_{{\rm He}} = 15.94$ cm$^{-3}$ and $n_{{\rm H}_2}
= 75.51$ cm$^{-3}$.
}\label{f01}
\end{figure}

We used the collisional dissociation rates of molecular hydrogen
provided in \cite{shapkang} for these three reactions.  We
also calculated the cooling function considering both the radiative
and the dissociative processes.   For the radiative cooling rate,
we used the  fit proposed by \cite{leppshull}, which considers
both the rotational and vibrational cooling from the two reactions,
H-H$_2$ and H$_2$-H$_2$ , in both high- and low-density regimes ($n
< n_{cr} \approx 10^4$ cm$^{-3}$).  The dissociative cooling function
was taken from \cite{shapkang}. In Fig.~ \ref{f01} we show the
different cooling functions: atomic (blue line) and molecular
dissociative (green line) and radiative cooling (red line) \footnote{In
order to calculate each one of these curves we considered the
initial values for the numerical densities for the different species
(atomic, ionic and molecular).}.

\begin{table*}
\caption{Jet models}                 
\centering                           
\begin{tabular}{c c c c c c c c c c c c}           
\hline\hline                         
Model & $v_{j,0}$ & A & $n_a$ & $n_j$ & $\eta$ & $T_a$ & $T_j$ & $\tau_P$ & $\theta$ & $N_{tot}$ & $\tau_e$  \\ [0.5ex] 

      & (km s$^{-1}$) &  & (cm$^{-3}$) & (cm$^{-3}$) & & (K) &  (K) &  (years) & ($^\circ$) &  &  (years)   \\ 
\hline
\hline
DR\_SS  &     212  &  0         & 100 & 100 &  1   & 100 & 1000     &     -   & - & -  & -     \\
DR   &        212  &  0.15      & 100 & 100 &  1   & 100 & 1000     &     -   & - & 4  & 5, 10, 20 and 50      \\
DR\_P  &      212  &  0.15      & 100 & 100 &  1   & 100 & 1000     &     200   & 6 & 4  & 5, 10, 20 and 50      \\
\hline
\hline
\end{tabular}
\tablefoot{
$v_{j,0}$ is the jet velocity in km s$^{-1}$; $A$ is the
amplitude of variation in the jet velocity; $n_a$ and $n_j$ are the
ambient medium and jet (numerical particle) density; $\eta = n_j/n_a$
is the jet to ambient medium density ratio; $T_a$ and $T_j$ are the
ambient medium and jet temperatures (in K); $\tau_P$ is the precession
period, $N_{tot}$ the number of different jet injection period, and
$\tau_e$ is the jet velocity variability period.}
\label{table1} 
\end{table*}

Together with H$_2$, CO and H$_2$O have long been known to play an
important role in shocked gas cooling \citep{holl,kaufman,flower}.
Detailed observational studies have confirmed that line cooling
from CO and H$_2$O can be as important as that from H$_2$ in protostellar
outflows \citep[see e.g.][]{nisini,busquet}.  Modeling of the
structure of outflow shocks may be significantly modified by the
inclusion of additional terms such as CO, H$_2$O, or even charged
grains \citep[see e.g.][]{flower}, all of which are not taken
into account here.  In the present work, we have not included either
the H$_2$O or the CO chemical networks or their related cooling
terms. This would represent an effort which is well beyond the state
of the art of 2D and 3D chemo-hydrodynamical codes such as {\sc
WALKIMYA-2D} \citep[][Rivera-Ortiz et al., in prep.]{cast18b}. 
However, we note that \cite{rs04} included equilibrium C and O
chemistry in their numerical scheme in order to calculate the CO,
OH, and H$_2$O abundances, as well as to estimate the cooling expected
from these molecules. They concluded that the mass--velocity
relationship is always shallower than the intensity--velocity
relationship, confirming previous results based only on H$_2$
\citep{downes} \footnote{The CO intensity--velocity is calculated
implicitly in \cite{downes} using an analytical prescription and
the local density, assuming that CO density is 10$^{-4}$ of the
H$_2$ density.}. For that reason, the present work focuses on the
entrained gas properties and we aim to revisit the H$_2$ mass--velocity relationships, whose properties can be accessed following
the present prescription.

Our computational domain is a  Cartesian 3D rectangular
box with the following dimensions:

\begin{equation}\label{sizebox}
(x,y,z) = (2, ~2, ~8)\times10^4 ~{\rm au} \, 
,\end{equation}

\noindent and the jet propagates along the $z-$ direction.

The Yguaz\'u-a is a multi-level binary adaptive grid code. Here
we use a five-level grid which has ($x$, $y$, $z$) = (256, 256, 1024)
cells in its high-resolution mode.  This gives a maximum resolution
of $\Delta x = \Delta y = \Delta z = 78.13$ au. The jet radius is
initially always given by $R_j = 391$ au or $\sim 5 \Delta x$.
The jet radius is therefore compatible with those adopted in previous
numerical simulations \citep{DR99, downes} as well as with estimates
for the HH jet radius \citep{rei00, rei02, podio2006}.

\subsection{Physical conditions}\label{simu}

Three cases were considered, which are summarized in Table \ref{table1}:

\begin{itemize}
\item model DR: an intermittent jet model for comparison with
previously published simulations  in the literature (DC03, Downes
\& Ray 1999);
\item  model DR\_SS: a steady state jet;
\item model DR\_P: an intermittent, precessing jet model.
\end{itemize}

With model DR\_P, we aim to investigate the properties of the L1157
outflow,  kinematical studies of which have revealed convincing evidence
of precession \citep{gueth1996, podio2016}. In this simulation (and
for model DR) we  assume that the jet velocity varies periodically
with time, according to:

\begin{equation}\label{pul}
v_j = v_{j,0} \cdot \bigg[ 1 + A\sum_{i=1}^{N_{tot}}{\rm sin} 
\bigg( \frac{2 \pi}{\tau_{e,i}} \cdot t \bigg)
\bigg]\,,
\end{equation}

\noindent where $v_{j,0}$ is the jet velocity, $A = \Delta v/v_{j,0}$
is the adopted amplitude variation for the jet velocity variation,
and $\tau_e$ is the variability period ($t$ is the time).  In our
time-varying models, $N_{tot} = 1$ or 4, and $5 \lesssim \tau_{e,i}
\lesssim 50$ years (see Table \ref{table1}).  We note that although
Eq. \ref{pul} has been used here in an attempt to reproduce the
model presented in DC03, the idea that Herbig-Haro objects can be
generically explained by successive internal knots promoted by a
sinusoidal jet velocity variability is well established
\citep[e.g.,][]{rei01}.  However, a detailed source modeling can
require a superposition of different sinusoidal terms, which have
been discussed by Castellanos-Ramírez, Raga \& Rodríguez-González
(2018; see also Bally 2016), indicating that a multimode jet
velocity variability may be important to explain the observed
morphology and kinematics in some sources.  For the precessing case
DR\_P, we adopted a precessing angle of $\theta = 6^{\circ}$ and a
precessing period of $\tau_P = 200$ years. The DR model has the
same parameters as the model presented in DC03.

In all models, we assume  solar elemental abundances for both ambient
and  jet material.  The ratio $n_{{\rm H}_2}/n_{\rm H} = 9$ (here
and after, $n_{\rm H} = n_{\rm HI} + n_{\rm HII}$) is initially
imposed for both the jet and the ambient medium \citep{DR99,
nisini10}.  The helium fraction per hydrogen nuclei is assumed to
be $n_{\rm He}/[n_{\rm H} +2n_{{\rm H}_2}] = 0.1$ With these choices,
the equation of state is calculated for a gas with a mean molecular
weight of $\mu = 2.23$ and $C_V = 2.25$.  The ionization fraction
of hydrogen in the jet is initially taken as $f_H = 0.01$ for $T_j
= 10^3$ K in agreement with the values  inferred from atomic line 
observations of HH jets in the optical \citep{podio2006}.

The ambient medium and jet parameters such as numerical density
$n$, temperature $T$, and jet velocity  are all given in Table
\ref{table1}, along with the jet precessional and
intermittence periods of the simulated models.

Observationally, the jet temperature and density determinations
span a wide range of values depending on the tracer used.  Optical
atomic line observations yield $T_j \sim 5\times10^3-2\times10^4$
K \citep{podio2006} while molecular line observations indicate lower
values of about  $10^3-3\times10^3$ K from  near-infrared H$_2$
rovibrational transitions \citep[e.g.,][]{caratti},  and $T_j \sim$
100-500 K from (submillimeter)  CO and SiO rotational
lines \citep{nisini07,lefloch15}.  In the optical, inferred jet
densities have values of $n_j \sim 10^3-10^4\cmmt$ , while (sub)millimeter
line observations yield high values, $n_j \gtrsim 10^5$ cm$^{-3}$.
This wide range of physical conditions reflects the intrinsic
complexity of the jet, which is often associated with internal shocks that
drive the formation of strong temperature and density gradients.
Adopting single initial values for temperature and
density  is most likely an oversimplified description of the jet physical
structure.  We note however that the initial jet temperature value
adopted in the simulations are consistent with those obtained from
jet molecular line observations  (H$_2$, SiO, CO).  The initial jet
density in our simulations is $100\cmmt$ (same as in DC03),  which is lower
than the values determined observationally. However,  it is
the jet-to-ambient density ratio which carries the most weight in modeling the
dynamical evolution of the outflow. This point was investigated in
detail by \cite{rs04b}. Based on their results,  we do
not expect significant differences  in the simulations when adopting
a higher density  for the jet, provided that the jet-to-ambient density
ratio is kept constant.

\subsection{Mass--velocity profiles}\label{mvprof}

\begin{figure}
\centering
\includegraphics[width=\columnwidth]{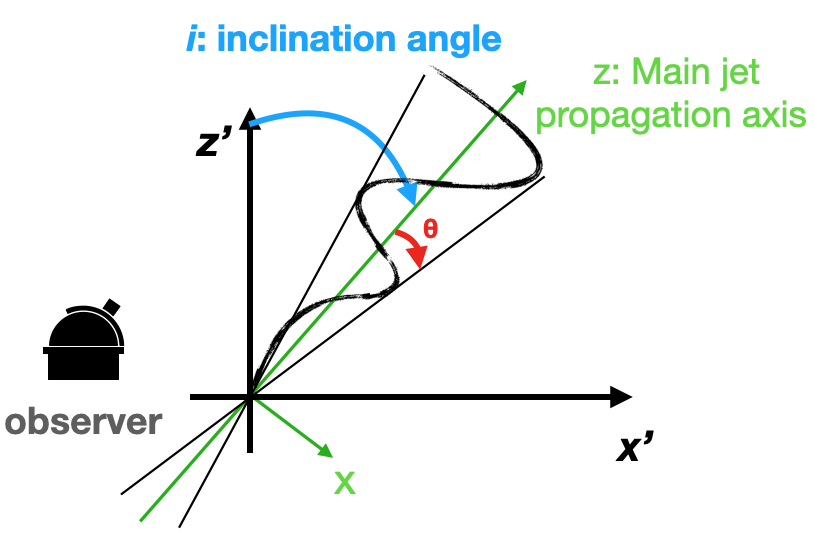}
\caption{Sketch of the geometry of the flow with respect to the
observer.  The jet propagates along the $z$-axis, which is inclined
by an angle $i$ with respect to the plane of the sky ($y^\prime -
z^\prime$).  In case of precession,  the jet draws a cone with a
half angle of $\theta$ with respect to the $z$-axis.}
\label{f02}
\end{figure}

Our primary diagnostics are the mass--velocity relationship for both
the total mass $m(v)$ and the molecular mass, $m_{{\rm H}_2}(v)$,
which are computed  for the whole computational domain or for a
given spatial region. In order to obtain the mass--velocity profiles,
we first compute the column density $N$ as the sum of particle
density along the line of sight per velocity interval:

\begin{equation}\label{colden}
N (y^{\prime},z^{\prime}) \bigg\vert_{v = v_{\rm CM}} = \sum n^{\prime} \Delta x^{\prime} ~\,,
\end{equation}

\noindent where

\begin{equation}\label{vrad}
v = v_x{\rm cos}i - v_z{\rm sin}i ~.
\end{equation}

\noindent In Equation (\ref{colden}),  $\Delta x^{\prime}$ is the
projection of the $x-$coordinate along the line of sight and
$n^{\prime}$ is the numerical particle density (total or molecular)
in the radial velocity range $(v - \Delta v/2) < v < (v + \Delta
v/2)$, where we set $\Delta v= 1\kms$.  In Equation (\ref{vrad}),
$i$ is the inclination angle with respect to the plane of the sky
(see Fig.~\ref{f02}), meaning that $v$ corresponds to the (observed)
radial velocity.

As the jet propagates, interaction with the ambient gas leads to
the formation of a mixed gas layer from jet and ambient material. In
order to estimate the relative contribution of the different regions
of the outflow, namely  the swept-up ambient medium and the mixed
(jet plus ambient medium) material, we tagged the jet
material with a passive scalar or {\it tracer} $j$. This passive
scalar is set to $j=0$ in the ambient medium and to $j=1$ in the
jet, and is advected by the flow. As the jet interacts with the
ambient medium and mixing occurs, the local value of the scalar represents the relative fraction of initially jet material within a
given cell. With this we introduce a parameter $j_{mix}$ to indicate
the level of mixing considered, such that $j \leq j_{mix}$. Thus,
material with $j_{mix}=0$ would correspond to ambient unmixed medium, that with
$j_{mix}=0.5$ would include material that has a jet fraction up to
$0.5$, and that with $j_{mix}=1$ will include all jet and ambient material.

\subsection{Comparison with previous work}\label{comparison}

We tested our numerical scheme against previous numerical
simulations published in the literature by running model DR (Table
\ref{table1}). The parameters of model DR were chosen to mimic model
G in \cite{DR99}, as well as the ``pulsed'' model discussed in
\cite{downes}.

Briefly, the DR model has jet ($n_j$) and ambient medium ($n_a$)
numerical particle density $n_j = n_a = 100$ cm$^{-3}$, jet
temperature $T_j = 1000$ K, ambient medium temperature $T_a
= 100$ K, molecular hydrogen to hydrogen numerical density
ratio $n_{{\rm H}_2}/n_{{\rm H}} = 9,$ and a helium to hydrogen
numerical density ratio of $n_{{\rm He}}/n_{\rm H} = 0.1$. The only
H$_2$ dissociation process considered in this model is  collisions
with H atoms, and the rate coefficient used here is given by:

\begin{equation}\label{kd1TR95}
k_{D,H-H_2} = 2\times 10^{-10}{\rm exp}(-55\,000/T) ~~ {\rm cm}^{3}{\rm s}^{-1}~\,
,\end{equation}

\noindent following \cite{TR95}. 

\begin{figure*}
\centering
\includegraphics[width=520pt]{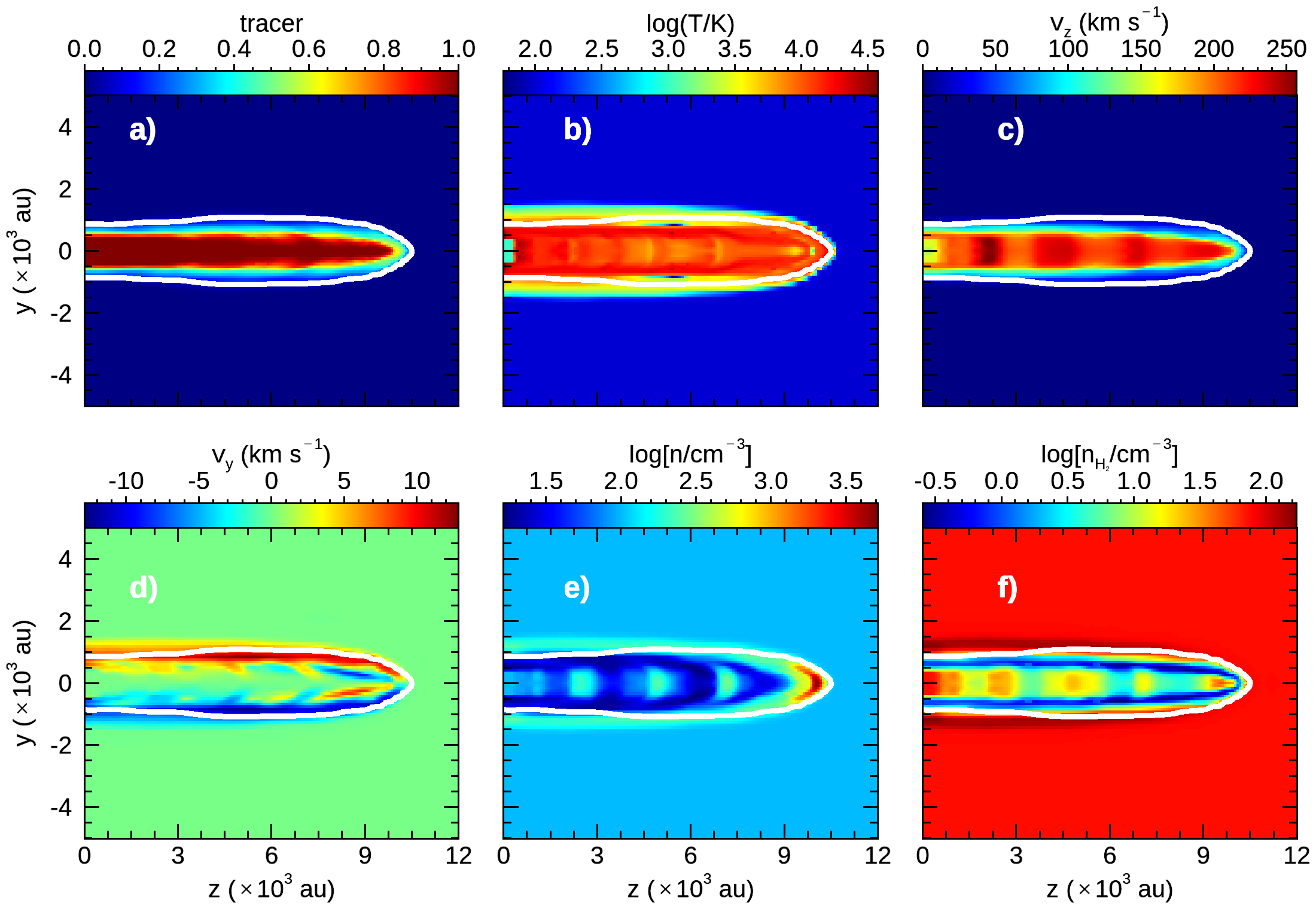}
\caption{ Model DR at $t$= $400\yr$. Distributions in the plane
$x$=0 of $(a)$ $j_{mix}$ (tracer), $(b)$ temperature, $(c)$ $v_z$,
$(d)$ $v_y$, $(e)$ total density $n,$ and ($f)$  molecular gas
density $n_{{\rm H}_2}$.  The white line in each panel separates
the region in the computational system filled by the ambient medium
---where the tracer is equal to zero--- from the mixing region and the
jet itself.}\label{f03}
\end{figure*}

Figure~\ref{f03} shows the results of model DR at $t$= $400\yr$.
Depicted in this figure are the midplane ($x = 0$) distribution
of the tracer (panel $a$), the temperature (panel $b$), the velocity
components along the $z-$axis (panel $c$) and the $y-$axis (panel
$d$), the total particle density $n$ (panel $e$), and the molecular
particle density $n_{{\rm H}_2}$ (panel $f$).  We plot a
white contour line that separates the original (quiescent and/or
disturbed) ambient medium where the tracer $j_{mix}$ is zero (the
region outside the contour line) from the inner jet  where $j_{mix}=
1$ (see Fig. \ref{f03}).

Figures~\ref{f03}($e)$ and \ref{f03}$(f)$ show the internal working surfaces   at $z$= 2500,
5000, and 7500 au, where both the total
density and  the molecular density increase.
The leading bow shock
is more pronounced in the density maps (panels $e$ and $f$).  Although
density enhancement is observed in the internal shocks (see
Fig.~\ref{f03}) as expected, neither the total density nor the
molecular density attain their maximum values at the internal
working surfaces. The region where the total density is maximum is
located at the head of the jet  and inside the contour line, indicating
that the jet and the ambient medium have already been mixed. It is
important to note that while the molecular density is higher at the
$external$ edges of the laterally expanding trails of the leading
bow shock, the total density peaks are near to the jet axis and the
apex of the leading bow shock.  This occurs because
strong shocks at the jet head dissociate the molecular hydrogen, while the shocks  are weaker at
the bow
shock trails and the molecular hydrogen  piles up without being dissociated.   We already anticipate
that the mass--velocity profiles extracted from our simulations
should present a high-velocity component  ($v \sim 100 \kms$) if
some mixing between the jet and the ambient medium material is
allowed.

\subsubsection{Comparison with Downes and Ray (1999)}

\begin{figure}
\centering
\includegraphics[width=180pt]{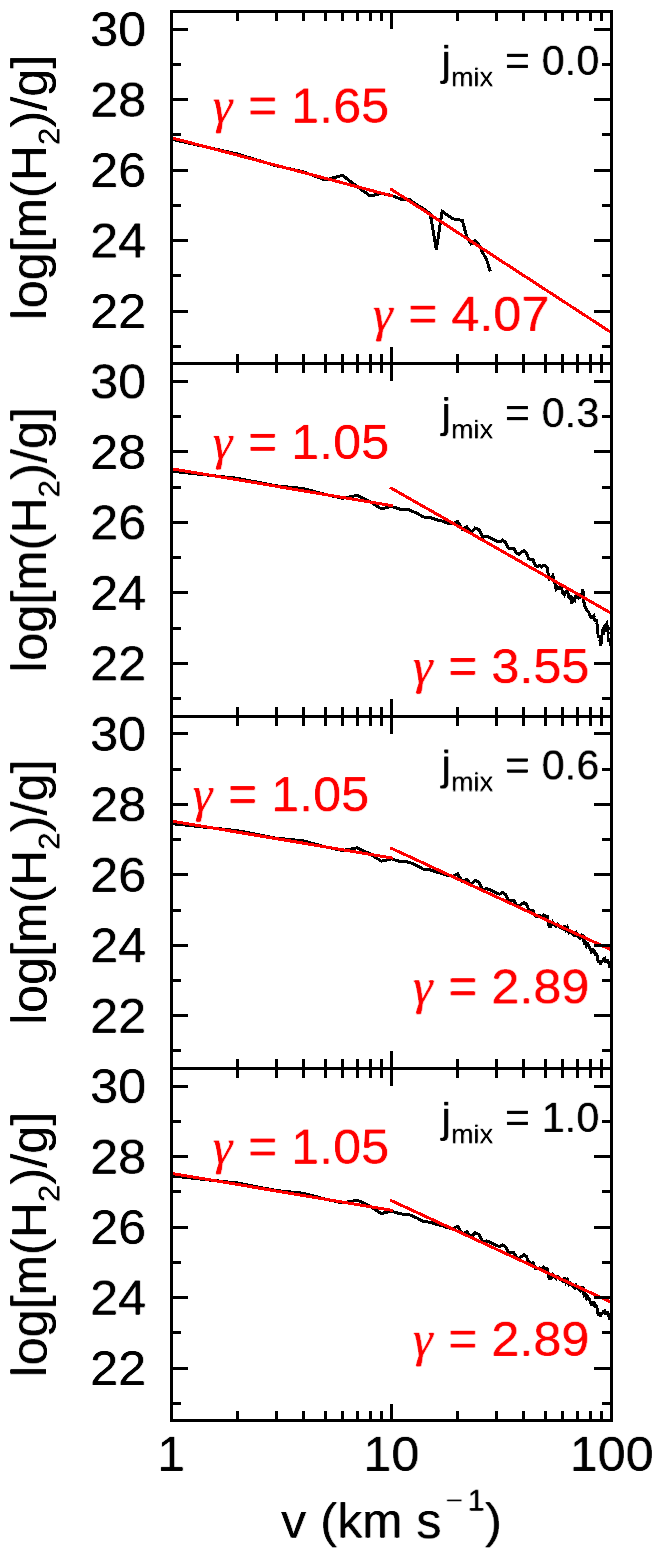}
\caption{Model DR at $t= 300\yr$ and an inclination angle of
$i=60^{\circ}$. Molecular mass--velocity relationship $m_{{\rm
H}_2}(v)$ (black) for the $ 0 \kms < v < 100 \kms$ radial velocity
range and the best-fitting power law $m(v) \propto v^{-\gamma}$
(red). The best-fitting index value $\gamma$ is shown inside each
panel for two intervals: $ 0 \kms < v < 10 \kms$ (left) and $
10 \kms < v < 100 \kms$ (right). The $j_{mix}$ parameter is also
indicated (from top to bottom, $j_{mix} =$ 0, 0.3, 0.6 and
1.0). The vertical axis displays the logarithm of the molecular mass 
(in g).}\label{f04}
\end{figure}

In this section, we compare the results of model DR (see Table
\ref{table1}) with the results of  \cite{DR99}. More precisely, we
compare the results of the mass--velocity distribution at $t$=
$300\yr$ and for an inclination angle of $i =
60^{\circ}$  with respect to the observer for  a direct comparison with Fig. 3 in \cite{DR99}.
In Fig.~\ref{f04} we report  the  molecular mass--velocity profile
obtained when integrating over a region defined by a minimum level
of mixing between the jet and ambient medium material from $j_{mix}$=
0 (ambient only; top panel) to $j_{mix} = 1.0$ (from ambient plus
jet; bottom panel).  We have superposed the best fit obtained
using a power-law $m(v)\propto v^{-\gamma}$  in red.  For the sake of clarity,
we have made adjustments for two distinct radial-velocity intervals:
$ 0 < v < 10$ km s$^{-1}$ and $ 10 < v < 100$ km s$^{-1}$.
Hereafter, $v$ is used to refer to the radial
velocity.

The mass--velocity relations in Fig.~\ref{f04}  are well described
by a broken power law. At $v \sim 10\kms$, the slope changes in all
cases from $j_{mix} = 0$ up to $j_{mix} = 1.0$.  As expected, the
slope in the low-velocity range is always shallower than in
the high-velocity range.  However, there is an important
and systematic effect in the slopes caused by the jet material
removal from the integration process that results  in an overall
steepening in the mass--velocity relation.

We can see in  Fig.~\ref{f04} that considering different levels of
jet content in the computation of the mass--velocity relation induces
a similar effect: in the high-velocity range ($v> 10 $ km s$^{-1}$),
the slopes are steeper for lower values of $j_{mix}$.  By contrast,
in the low-velocity range ($v \le 10~\kms$) the slopes barely vary
with $j_{mix}$.  We interpret these facts  as a consequence of
effect of mass addition in a given velocity channel, when we go
deeper into the mixing layer.  In the low-velocity range the profile
is dominated by the material of ambient origin, which can be either
swept-up gas by the jet-driven bow-shock or entrained gas that
barely interacted with the jet.  The contribution of the jet and/or
ambient interacting material becomes increasingly apparent  when
considering increasing $j_{mix}$ values in the high-velocity range.

In their simulation run, \cite{DR99} obtained a mass--velocity
distribution which was best fitted with a power law of index $\gamma
= 2.98$   in the range $10 < v < 100$ km s$^{-1}$  (see
their Fig.~3).  In our simulation run with the same set of parameters
with Yguazú-a, we obtain  a similar mass--velocity relationship,
which can be fitted by a  power law (Fig. \ref{f04}). However, we
find that the power-law index depends on the degree of mixing between
the jet and the entrained ambient material, i.e., with the value of
$j_{mix}$. For the  {\em swept-up} (unmixed)  molecular material
($j_{mix} = 0$),  the mass--velocity relationship is best fitted by
a power law of index $\gamma = 4.07$, which is steeper than the value found
by \cite{DR99}.  However, if we take into account the contribution
of mixed ambient and jet material  (e.g., $j_{mix} = 0.6$) to the
mass--velocity relationship, we obtain a best fit with a power law
with a shallower index of $\gamma$= 2.89,  which is very close to that found by \cite{DR99}.
This results  also holds when we take into account {all} the
material along the line of sight while assuming full mixing between the
jet and the ambient material ($j_{mix} = 1.0$).

We conclude that our simulations are in very good agreement with
those of  \cite{DR99}. We can retrieve their  results with excellent accuracy in the limit of large mixing degree  ($j_{mix}\ge
0.6$). Our results suggest that the jet contribution must also be
taken into account when studying the mass--velocity distribution.

\subsubsection{Comparison with Downes and Cabrit (2003)}

Using a numerical code very similar to that of \cite{DR99} and
with similar initial conditions, DC03  further investigated  the
mass--velocity and intensity--velocity relations in the CO J=2--1 and
H$_2$ S(1) v=1--0 lines for jet-driven molecular outflows.

\begin{figure}
\centering
\includegraphics[width=\columnwidth]{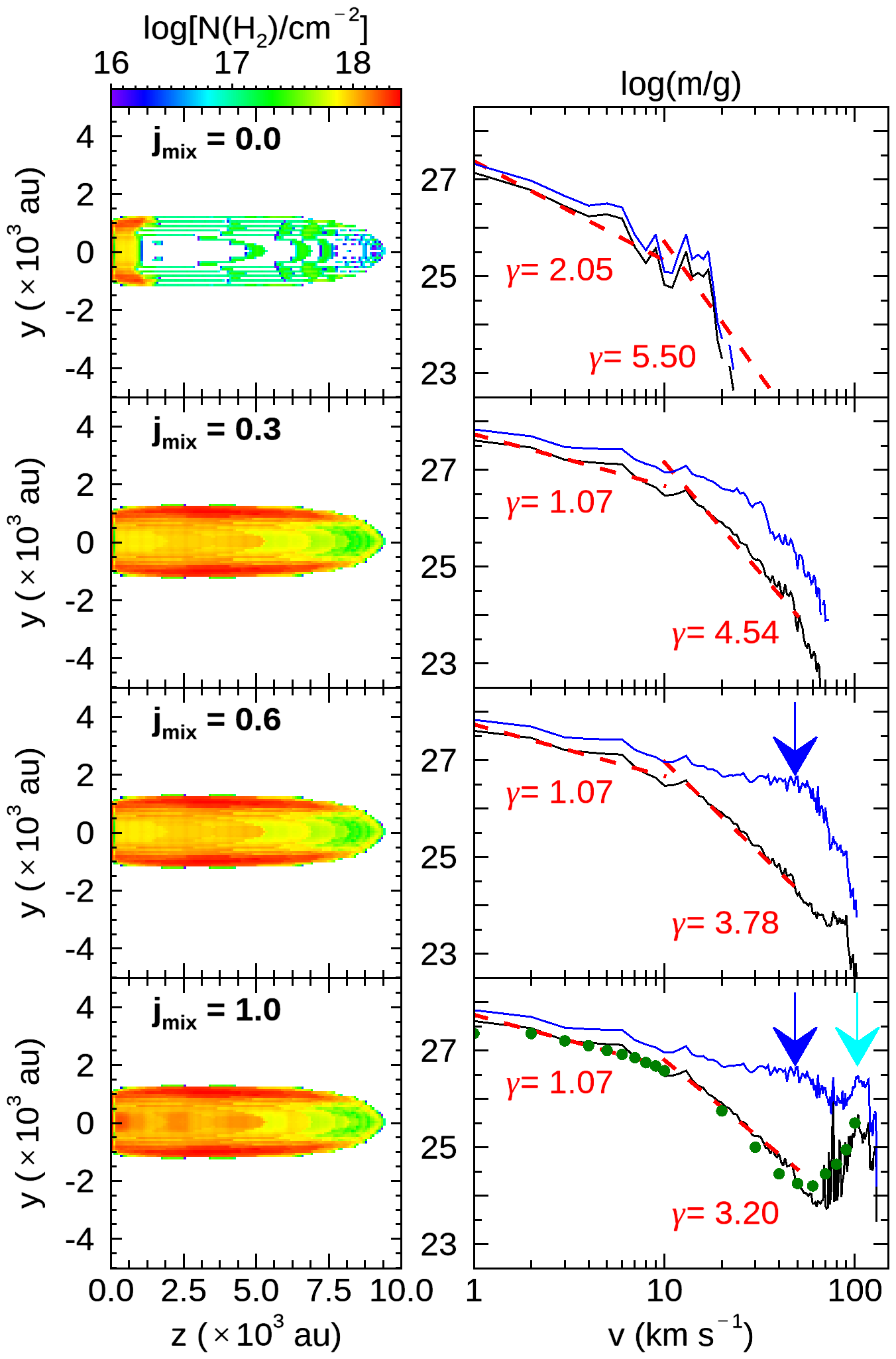}
\caption{Model DR at  $t = 400$ yr and an inclination angle $i$=
$30^{\circ}$. The results are presented for four different  values
of the jet and ambient gas mixing degree $j_{mix}$ (from top to bottom):
0.0, 0.3, 0.6, 1.0.  {\em (left)} Maps of H$_2$ column density
obtained by integration over the velocity range, between $+1$ and
$+150\kms$.  {\em (right)} Mass--velocity relationship obtained for
the molecular gas (black) and the total gas (blue).  In the bottom
panel ($j_{mix}=1.0$), the green dots correspond to  the H$_2$
swept-up mass--velocity relationship obtained by DC03 (see their
Fig.~2).  We show the best-fitting power
laws $m(v)\propto v^{-\gamma}$  to the velocity intervals $v \le
10 \kms$ and $10 < v \lesssim 20 \kms$ superposed in red dashed lines.  The power-law index
$\gamma$ is given in each panel.  The vertical arrows indicate the
position of bumps in the total mass--velocity profile (see  text).}
\label{f05}
\end{figure}

In Fig.~\ref{f05} we show maps of H$_2$ molecular column density
(left panels) and the mass--velocity relationships  (right panels):
the molecular $m_{{\rm H}_2}(v)$ (black lines) and the total
mass--velocity relations $m(v)$ (blue lines).  For the sake of direct
comparison with the results presented by DC03 (see their Fig.~2),
the molecular mass $m_{{\rm H}_2}(v)$ and mass $m(v)$ velocity
relationships have been extracted from the DR model at $t = 400$
yr, considering an inclination angle of $i=30^{\circ}$ and different
values of the jet and ambient gas
mixing ratio $j_{mix}$= 0, 0.3, 0.6, 1.0.
In order to obtain the molecular mass distribution of the outflow--jet
system,
we integrated over the full range of velocity, between
$+0$ and $+150 \kms$, and excluding the emission of the
quiescent gas at rest velocity ($v$=0). By doing so, the contrast
of the image is enhanced, easily revealing the geometry of the
cavity and the presence of the internal working surfaces produced
by the jet.

Comparison of the mass--velocity relationships obtained for different
values of $j_{mix}$ in Fig.~\ref{f05} with respect to the swept-up
gas (top panel) shows  the presence of gas accelerated at $v >
10\kms$ already for $j_{mix} = 0.3$ and $j_{mix} =0.6$, which
results in a shallower mass--velocity distribution.  This change of
slope  is mainly caused by the contribution of the massive gas clump
that develops near the apex of the bow shock inside the mixing
layer. This effect can be seen  in  Fig.~\ref{f03}(e).

The total and molecular mass--velocity relationships are very close
to each other in the swept-up gas (top panel in  Fig.~\ref{f05}),
which implies that the molecular gas dissociation can be neglected
in the local gas acceleration (entrainment) process. When taking
into account the jet--ambient gas mixing,  the slope of the total
mass--velocity relationship (blue in Fig.~\ref{f05}) starts to depart
from the molecular mass--velocity slope (black) for $j_{mix} = 0.3$
case.  The difference increases with increasing velocity and
increasing $j_{mix}$ (jet-ambient mixing ratio) values.  The change
of slope between $j_{mix}$=0.3 and $j_{mix}$= 0.6 occurs near $v
\sim 50\kms$. We note that this velocity  coincides with the projected
velocity of the high-density gas concentrated at the jet head.
This region can be seen in Fig.~\ref{f03} (e) near $z = 9.5\times
10^3$ au. The velocity component of this dense component along
the jet axis is $v_z \approx 100\kms$, which corresponds to a
projected (radial) velocity of  $50 \kms$.  The increase of the
mass at high velocities (in blue in Fig. \ref{f05}) with increasing
$j_{mix}$ values is essentially unnoticed in the molecular mass--velocity distribution (in black in Fig. \ref{f05}). This  is
consistent with efficient H$_2$ dissociation in the shocks at the
jet head.

In the case of full jet--ambient gas mixing ($j_{mix}=1$), a  second
bump is detected in the mass--velocity distribution at $v \approx
100\kms$  (Fig. \ref{f05}), which is the signature of the internal knots
formed as a result of jet variability.  This second bump is present
in both the molecular (black lines) and the total mass (blue line)
velocity distributions. Again, this is illustrated  in Fig.~\ref{f03},
where both total and molecular densities appear to be  enhanced
behind internal shocks  at a  velocity consistent with the projected
velocities of the second bump.  We emphasize that this second bump
can only be seen if we consider the total jet material in the
calculation ($j_{mix} = 1.0$), while the first bump starts to appear
at moderate values of $j_{mix}$.

In the bottom right panel of Fig.~\ref{f05} we have superimposed
the results of the \cite{downes} simulation with green bullets.  As
we can see, the match between the swept-up H$_2$ --velocity
distribution of these latter authors and the results of our simulations is very close.
However, it should be noted  that DC03 claimed to have obtained
the molecular mass--velocity relation for the swept-up H$_2$, hence
excluding the jet material from the computation (see their Fig.~2)
. While DC03 claim that the jet has not been considered in their
computation of the mass, we were  only able to reproduce their profile
when taking into account the jet contribution in the computation. The
presence of a high-velocity bump in the H$_2$ mass--velocity
distribution  is evidence that the jet has been considered in
the integration procedure, as discussed above.

To summarize,  we carried out a detailed comparison of the
Yguazu-a code results with numerical simulations presented by Downes
and Ray (1999) and DC03. We obtained excellent quantitative
agreement in both cases. This bolsters our confidence in the use of this approach and in the results for the mass--velocity distributions
produced by our code.

\section{Results}\label{results}

\subsection{Steady state}\label{SSmodel}

\begin{figure}
\centering
\includegraphics[width=\columnwidth]{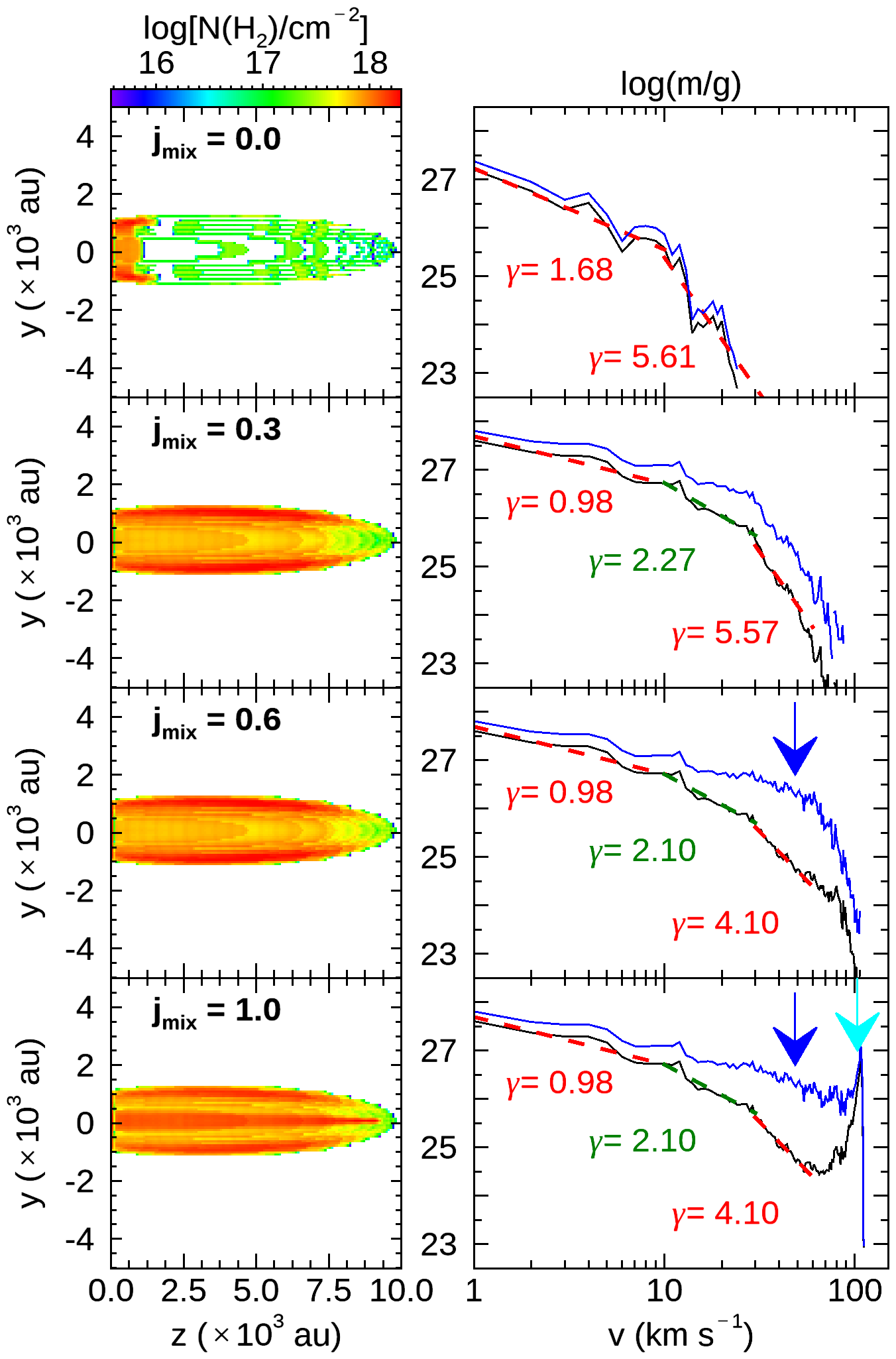}
\caption{Model DR\_SS at  $t = 400$ yr and an inclination angle
$i$= $30^{\circ}$. The results are presented for four different
values of the jet--ambient gas mixing degree $j_{mix}$ (from top to
bottom):  0.0, 0.3, 0.6, 1.0.  {\em (left)} Maps of H$_2$ column
density obtained by integration over the velocity range between 1
and $+150\kms$.  {\em (right)} Mass--velocity relationship obtained
for the molecular gas  (black) and the total gas  (blue).  The
best-fitting power laws for the three  velocity intervals:
$v \le 10 \kms$ (left; dashed red), 10 km s$^{-1}$  $< v <$
30 km s$^{-1}$ (middle; dashed green) and 30 km s$^{-1}$  $< v <$
60 km s$^{-1}$ (right; dashed red) are drawn. The index $\gamma$
for each one of these intervals are shown inside each
panel. The vertical arrows indicate the position of bumps in the
total mass--velocity profile (see text).}\label{f06}
\end{figure}

We first present the results of model DR\_SS, which simulates the
propagation of a nonprecessing jet under steady-state (SS) conditions.
The parameters of the simulation are summarized in Table \ref{table1}.
Figure~\ref{f06} shows the \htwo\ column density spatial
distribution, and the  total and molecular mass--velocity distributions
obtained for different values of the jet-ambient gas mixing ratio
$j_{mix}$= 0, 0.3, 0.6, and 1.0 at $t = 400$ yr and for an inclination
angle of $i = 30^\circ$.

The main difference between model DR\_SS and the simulation discussed
above in  Sect. \ref{numeric} lies in the SS assumption, that is,
the absence of intermittency in the mass-ejection process.  Many
similarities are therefore observed  when comparing the mass--velocity
relationships obtained in both simulations at a common age of 400
yr, which are presented in Figs. \ref{f05}-\ref{f06}. The main
similarities can be summarized as follows:

\begin{itemize}

\item The total mass and the molecular mass--velocity relations
associated with the swept-up gas ($j_{mix} = 0$) are very similar, 
and are separated by only a small and constant vertical offset over the whole 
velocity range ($\lesssim 20 \kms$).

\item The gap  between the total mass and the molecular mass becomes
more evident with increasing values of $j_{mix}$ and increasing
velocity. For the case $j_{mix} = 0.3$, the curves end up at $v
\simeq 90 \kms$ showing a difference in mass (total versus
molecular) of about two orders of magnitude.

\item In the $j_{mix} = 0.6$ case,  a  bump is detected  in the
total mass--velocity relationship at $v \simeq 50 \kms$. The lack
of detection in the molecular mass--velocity distribution suggests
it is mostly of atomic origin and that it traces the signature  of
material locally accumulated behind the jet head as a result of
shock compression.

\item A second bump of mainly atomic jet  material is detected at
high velocity, namely $v \sim 100 \kms$ (see panel $j_{mix} = 1.0$ in
Figs.~\ref{f05}-\ref{f06}).

\end{itemize}

We therefore conclude that the internal knots that form as a result
of the intermittency of the mass-ejection process do not fundamentally
alter the mass--velocity relationship obtained in a continuous
SS ejection process.  However, some differences are seen between the two simulations when comparing the relative contributions of the molecular and
atomic material.

The high-velocity bump detected at about $100\kms$  is related to
the internal knots  in the case of DR, whereas it traces the jet
material in the SS model.  Though mainly atomic, the bump contains
a significant fraction of molecular material.  The gap between the
total (atomic+molecular) and the molecular mass is wider in the
case of intermittent ejection (model DR; Fig.~\ref{f05}) than in
the SS regime (Fig.~\ref{f06}). Indeed,  H$_2$  dissociation  is
expected to occur in the multiple internal shock knots produced in
the pulsating model; by comparison, in the SS model, only the leading
bow shock is expected to dissociate $\htwo$. This is also well
illustrated by panels (e) and (f) in Fig. \ref{f03}.

The second bump detected at $v \sim 100 \kms$ in the DR model (see
Fig. \ref{f05}) is fully developed.  This indicates that the jet
material reaches a higher velocity in the SS regime, $\geq 100\kms$.
As can be seen in Fig~\ref{f06}, in the SS regime, the jet reaches
a velocity close to the maximum expected value, $v = 212\cdot{\rm
sin}30^\circ = 106 \kms$.  On the contrary, in a pulsating jet, the
formation of internal shocks results in a deceleration of the jet
material all along its axis.

We determined the best-fitting power law  to
the molecular mass--velocity relationship  for model DR\_SS in the three velocity
intervals: $v < 10$ km s$^{-1}$, 10 $\kms < v < 30 \kms$ and
30 km s$^{-1}$ $ < v <$ 60 km s$^{-1}$.  The results are
reported in Fig. \ref{f06}, where we show the results of model
DR\_SS  at $t = 400\yr$ under an inclination $i = 30^\circ$ with
respect to the plane of the sky (see Fig.  \ref{f02}). The slopes
of the molecular mass--velocity relationship decrease as $j_{mix}$
increases. In the low-velocity range, the behavior  of the
relationship is very similar to that obtained for a pulsating jet
(model DR; Fig.~\ref{f05}).  However, for the $v > 10 \kms$
and $j_{mix} \ge 0.3$, there is a region at intermediate radial
velocities (10 $\kms$ $< v <$ 30 $\kms$) with a moderate slope
($\gamma \sim 2$), followed by a region of a steeper slope ($\gamma
\sim 4-5$) at high velocities ($v > 30 \kms$). Although a direct
comparison with the slopes of the DR model (Figure \ref{f05}) at
intermediate-to-high velocity is impossible because in that case the
whole profile (from 10 $\kms < v < 60 \kms$) seems to be well
described by a single power law, we can roughly estimate a mean
$\gamma$ value for DR\_SS model using the two $\gamma$ values obtained
for $v > 10 \kms$, and we obtain $\gamma = 3.92$, 3.1, and 3.1
for $j_{mix} = 0.3$, 0.6, and 1.0, respectively. This suggests that
the molecular mass--velocity profiles at intermediate-to-high radial
velocities for the SS model tend to be shallower in
comparison with those obtained for the pulsating model, indicating
a high molecular mass content, a result that is consistent with the
scenario described in the previous paragraphs.

\subsection{Precession}

\begin{figure}
\centering
\includegraphics[width=\columnwidth]{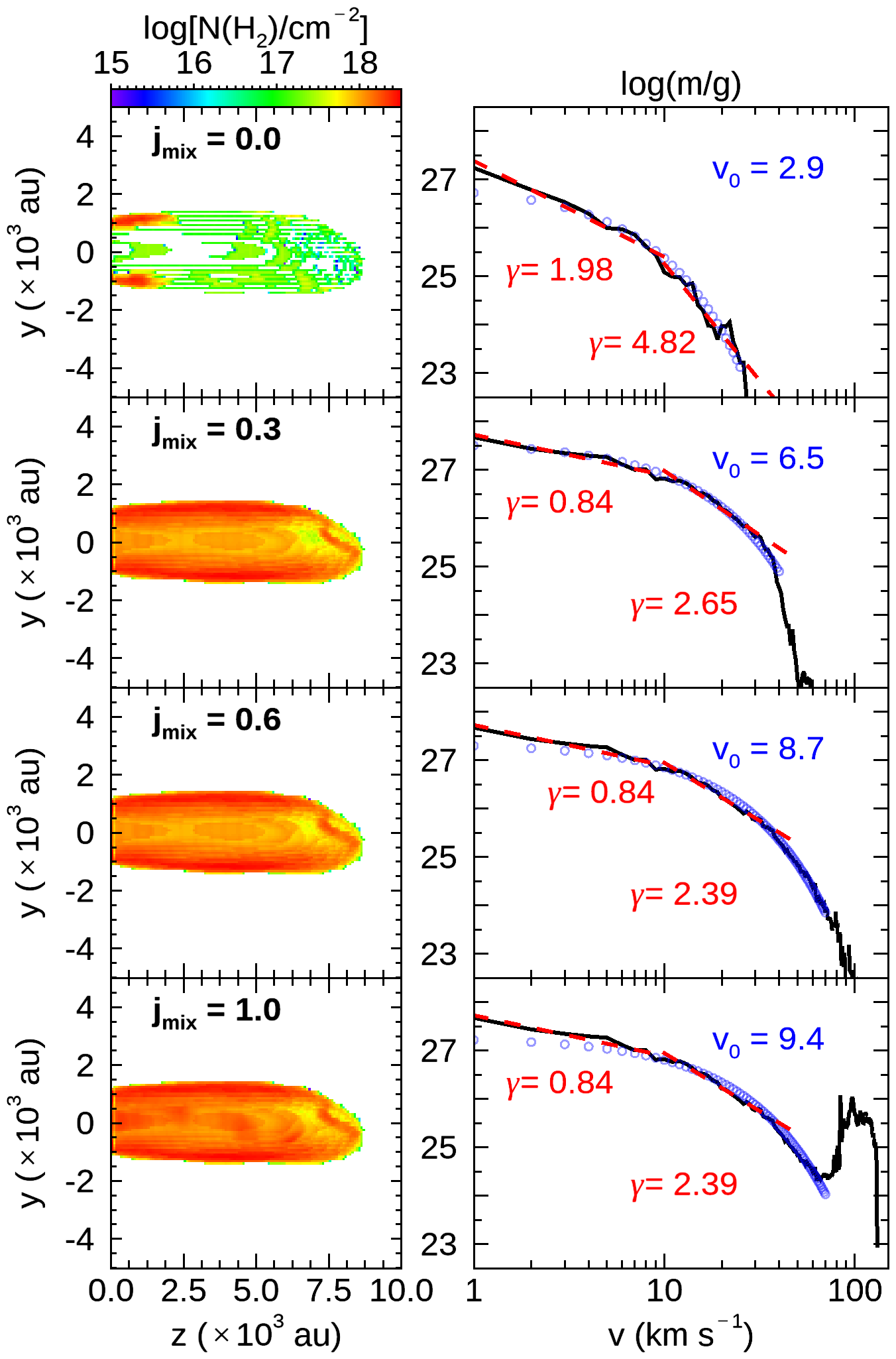}
\caption{Model DR\_P at  $t = 400$ yr and an inclination angle $i$=
$30^{\circ}$.  Results are presented for four different  values
of the jet--ambient gas mixing degree $j_{mix}$ (from top to bottom):
0.0, 0.3, 0.6, 1.0.  {\em (left)} Maps of H$_2$ column density
obtained by integration over the velocity range between 1 and
$+150\kms$.  {\em (right)} Molecular mass--velocity relationship (black).
The best-fitting power laws for the two  velocity intervals: $v \le
10 \kms$ (left) and $10 < v \lesssim 20 \kms$ are drawn in dashed
red. The index $\gamma$ is shown inside the panel. The exponential best
fit is drawn in blue and the exponent $v_0$ is given for each
$j_{mix}$ value.}
\label{f07}
\end{figure}

As many jets display hints of precession
\citep[e.g.,][]{gueth1996,gueth1998,podio2016,santangelo}, we 
investigated the possible  impact of this phenomenon on the   mass--velocity relationship
by running model DR\_P (see Table \ref{table1}). We modeled precession
with an angle $\theta = 6^\circ$ around the jet axis and a  period
of $\tau = 200$ years.  The column density maps and the total and
molecular mass--velocity distributions at $t=400$ yr and $i =
30^\circ$, and different values of the jet--ambient gas
mixing ratio
$j_{mix}$  are presented in Fig. \ref{f07}.

The cavity created by the precessing jet appears  broader close to
the apex when compared with unprecesssing models, either intermittent
(model DR; Fig.~\ref{f05}) or  SS (model DR\_SS;
Fig.~\ref{f06}).  Interestingly, both the molecular and total
mass--velocity distributions are very similar to those obtained in
the case of a pulsating, nonprecessing jet  (Fig.~\ref{f05}) and a SS
jet ( Fig.~\ref{f06}).  The same two bumps at $v\sim 50 \kms$ and
$v\sim 100 \kms$ are also detected  in the total mass--velocity
distribution (not shown in Fig. \ref{f07}).
In summary, the mass--velocity relationship is not significantly
altered by jet precession and is very similar to that of nonprecessing,
eventually pulsating jets.

\subsection{Exponential fitting approach}

\begin{figure*}
\centering
\includegraphics[width=520pt]{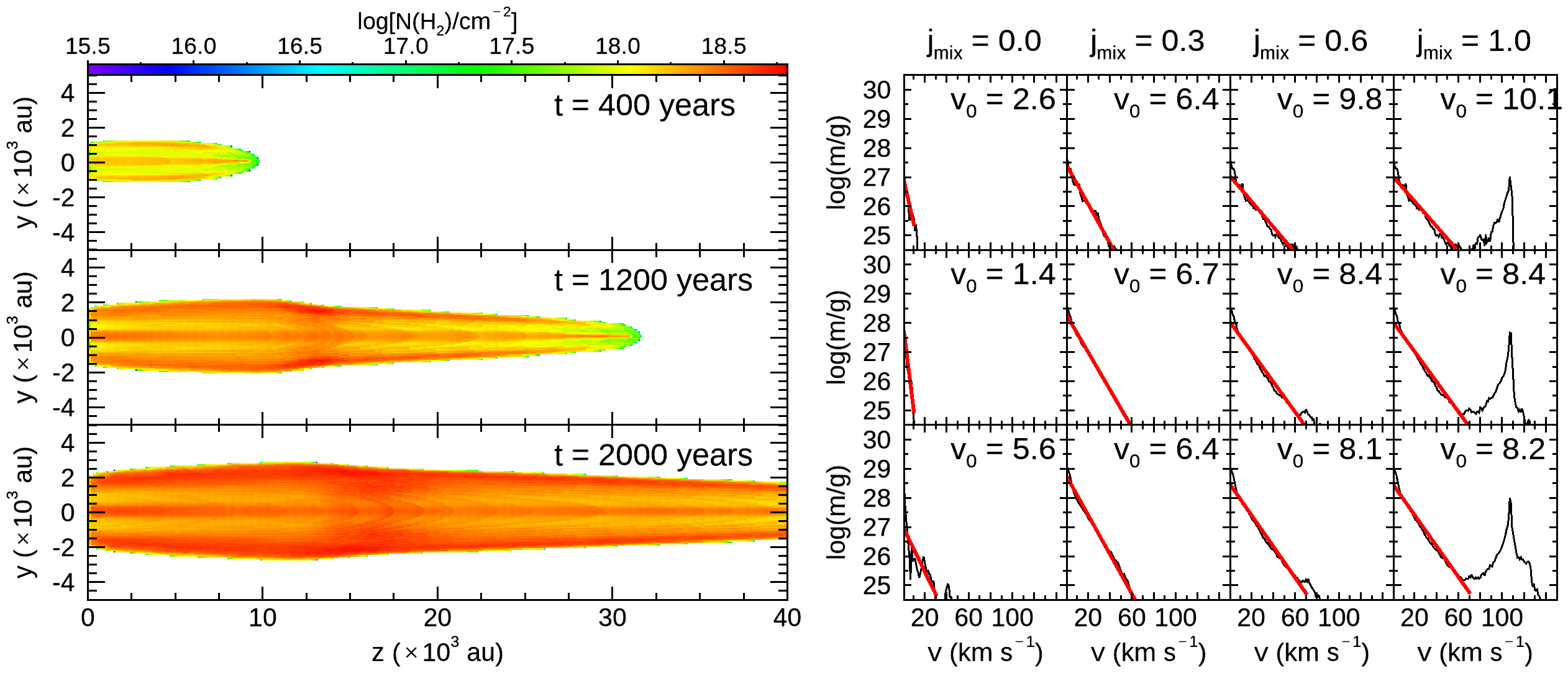}
\caption{Model DR\_SS  at $t$ = 400, 1200 and 2000 years (from top
to bottom) and for an inclination angle $i = 30^{\circ}$.  {\it
(Left)} Distribution of molecular gas column density. 
{\it (Right)} The mass--velocity relationship is depicted
in black for each value of the $j_{mix}$ parameter, from $j_{mix}
= 0.0$ (first column) to $j_{mix} = 1.0$ (fourth column).  The
profiles were calculated considering the whole computational domain.
The exponential best fit is drawn in red and the exponent $v_0$ is
given for each $j_{mix}$ value.}
\label{f08}
\end{figure*}

Above, we compare and analyze the results of our simulations
by modeling the mass--velocity distribution with a power law, $m(v)
\propto v^{-\gamma}$. However, observational work on several molecular
outflows by \cite{Lefloch2012} suggests that it could be possible
to adopt another fitting, namely an exponential law $m(v)\propto
\exp(-v/v_0)$. Based on our numerical simulations, we assessed the
validity of this approach.

\begin{figure*}
\centering
\includegraphics[width=520pt]{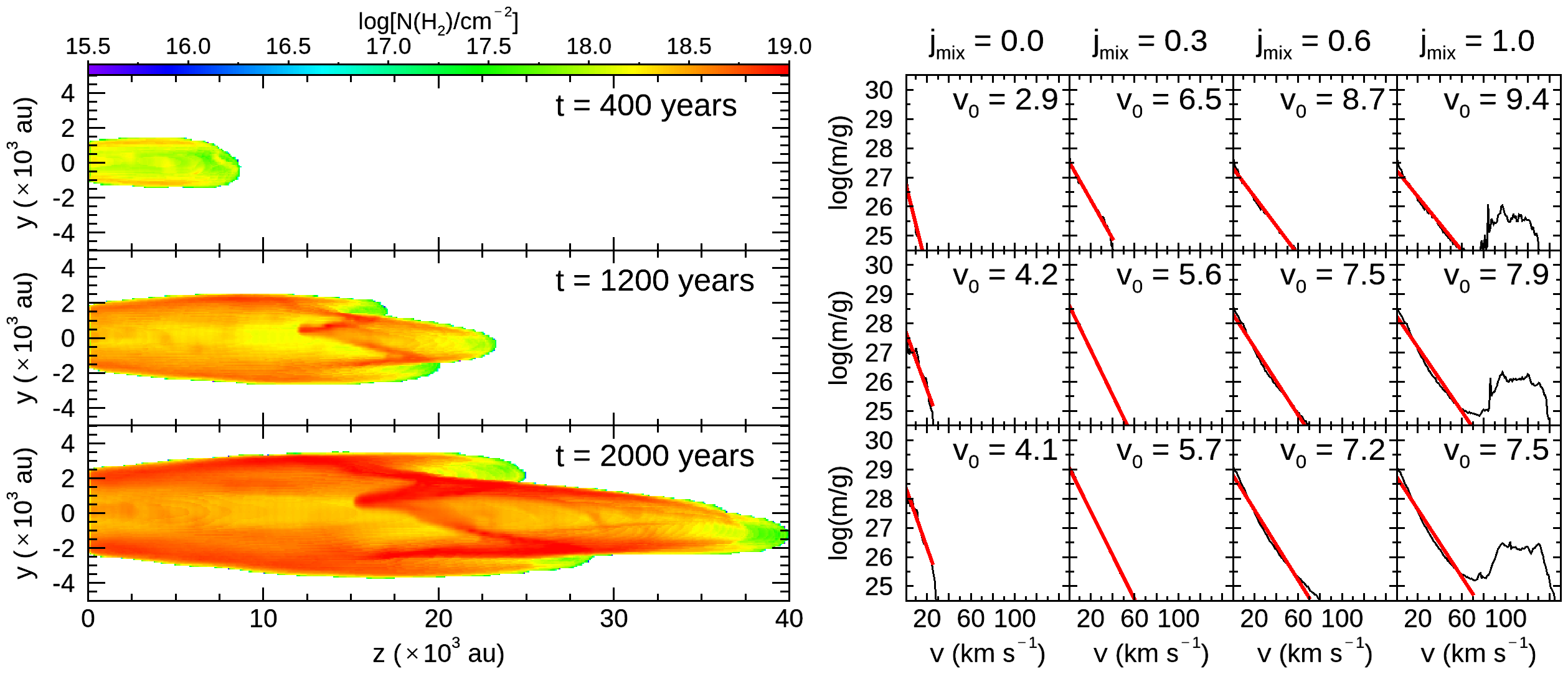}
\caption{Model DR\_P  at $t$ = 400, 1200, and 2000 yr and for an
inclination angle of $i = 30^{\circ}$.  The results are presented for
four different  values of the jet--ambient gas mixing degree $j_{mix}$
(from top to bottom): 0.0, 0.3, 0.6, 1.0.  {\em (left)} Maps of
H$_2$ column density obtained by integration over the velocity
range between 1 and $+150\kms$.  {\em (right)} Mass--velocity
relationship obtained for the molecular gas (black) and the fitted
curve (red).  The exponent $v_0$ is given for each $j_{mix}$
value.}\label{f09}
\end{figure*}

We show the results of the  fitting procedure for model DR\_SS (Fig.
\ref{f08}) and model DR\_P (Fig.~\ref{f09}) at the different times
$t$ = 400, 1200, and $2000\yr,$ and, as in the preceding analyses, for four
values of  the jet--ambient gas mixing ratio $j_{mix}$= 0.0, 0,3,
0.6, 1.0. We adopted  an inclination angle of $i= 30^{\circ}$.
We first consider the $\htwo$ column density distribution and the
molecular mass--velocity relationships  resulting from the propagation
of a steady-state jet (model DR\_SS) . The value of the  coefficient
$v_0$ is indicated in each panel.

Our first result is that it is indeed possible to obtain a very good fit between the mass--velocity relationship and an exponential
function $m(v) \propto {\rm exp}(-v/v_0)$ from  early
($400\yr$) to late ($2000\yr$) computational times (see
Figs.~\ref{f08}--\ref{f09}), although the  shallower $\gamma$
index in a power-law fitting translates into a higher $v_0$ value.
The value of $v_0$ therefore increases with the jet--ambient gas
mixing ratio.  Our simulations for DR\_SS (Fig~\ref{f08}) and
DR\_P (Fig.~\ref{f09}) show that higher values of $v_0$ are found
at earlier times ($400\yr$). Also,  the best-fitting values of $v_0$
for both models DR\_SS (Fig.~\ref{f08}) and DR\_P (Fig.~\ref{f09})
under the same inclination angle are similar at the different times
in the simulations, at 400, 1200, and $2000\yr$.  In other words,
regardless of the age of the system, once a minimum level of
mixing occurs between the jet and the ambient medium gas ($j_{mix}
\ge 0.3$), the mass--velocity relationship and the value $v_0$ are
almost unchanged.  However,  one can observe a small decrease of
$v_0$ as time increases.

Close inspection of the mass--velocity relationships reveals that
the exponential fitting  (red curves) provides excellent
solutions over the whole velocity range for $j_{mix}$ values of
0.3. For higher values of $j_{mix}$, the high-velocity range of the
distribution $v\geq 10\kms$ is still accurately fitted, with a
higher value  of $v_0$, as can be seen in Fig.~\ref{f09} (e.g., panels
$j_{mix}$=0.6).  This expresses the fact that the jet contribution
tends to make the mass--velocity distribution shallower,  as discussed
above.  On the other hand, the slopes in the  intermediate velocity
range depend on the mixing level, and we identify two mechanisms
that may explain this: the lack of material for $v \gtrsim 10$ km
s$^{-1}$ in the case of swept-up H$_2$ profiles (case  $j_{mix} =
0.0$) and the presence of the jet as a bump at $v \simeq 40$ km
s$^{-1}$ for $j_{mix} = 1.0$.  Between these two extremes we find great variability.  We note that some residual emission is
found in the low-velocity ($1 < v < 20\kms$)  range of the distribution,
which can be fitted by two exponential functions. This point is
addressed in more detail in the following paragraph.

\subsection{From large- to small-scale}\label{local}

The results that we have discussed so far refer to the global mass--velocity
relationship  computed over the whole computational domain.  However,
the possibility of a local exponential fitting was reported by
\cite{Lefloch2012} thanks to CO multi-line observations of the
shock position B1 in the southern lobe of the L1157 outflow. These latter
authors noticed that similar spectral signatures in the CO $J$=3--2
line were also detected at various other positions along the cavity
walls of the L1157 outflow. This leads us to speculate that these
spectral signatures could be a local property of the outfowing gas,
and not only a global property, as has been considered until
now.  The goal of this section is to explore this point further.

\begin{figure}
\centering
\includegraphics[width=\columnwidth]{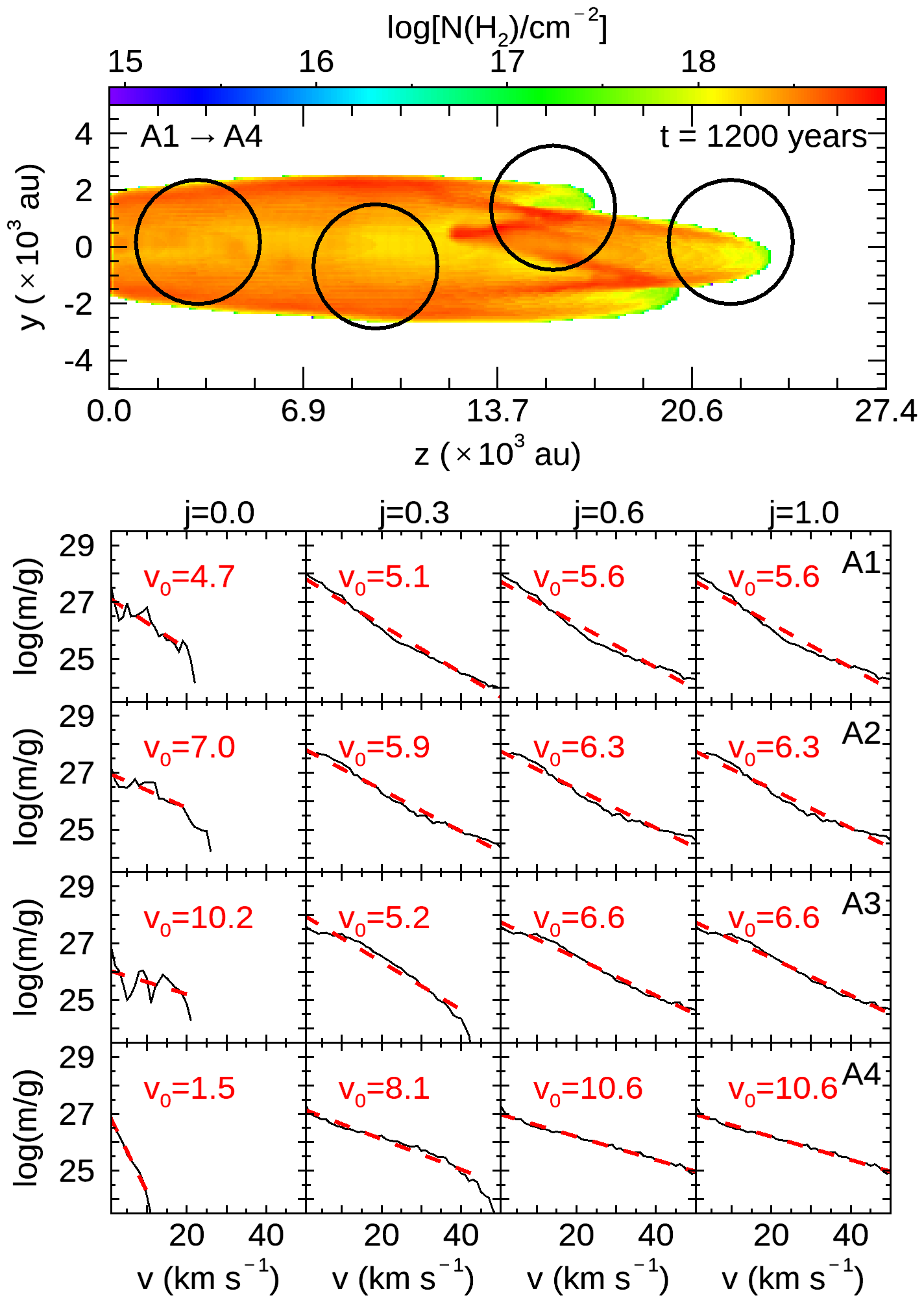} \caption{Model
DR\_P at $t$=$1200\yr$ under an inclination of $30^{\circ}$.  {\em
(top)} Map of H$_2$ column density  for $j_{mix} = 1.0$. The location
of the apertures A1-A4 used to extract the mass--velocity relationships
are drawn with black circles.  {\em (bottom)} Mass--velocity relationship
averaged over the circular apertures A1 to A4. The best exponential
fits are drawn in dashed red. The value of $v_0$ is given for each
value of jet--ambient material mixing ratio  $j_{mix}$= 0, 0.3, 0.6,
1.0.  }\label{f10}
\end{figure}

In order to investigate  the local behavior of the mass--velocity
relationship, we considered the simulation of the outflowing
gas in model DR\_P at $t= 1200\yr$,  with an inclination angle of
$i = 30^{\circ}$ and $j_{mix}= 1.0$. We selected four positions
in the  outflow, labeled  A1 to A4 and marked with black
circles in the map of molecular gas column density in Fig.~\ref{f10}.
While positions A1 and A2 are located inside the outflow cavity
with A1 close to the jet main axis, positions A3 and A4 are located
at shocked positions at the interface between the outflow and the
ambient gas.  We computed the molecular mass--velocity
relationships  over circular areas of $\sim$ 5000 au diameter at
the four positions. The relationships and the best fits are displayed
in the bottom panel of Fig.~\ref{f10}.

We verified that similar trends and results are found if we consider
an aperture larger than 5000 au.  However, profiles of the mass--velocity
distribution become irregular ("noisy") when the aperture size can
no longer be considered large enough in comparison with
the numerical
resolution of the  simulation.

Our first finding is that, at all four positions, the mass--velocity
distribution can be well approximated by an exponential fit
$m(v)\propto \exp(-v/v_0)$. Hence, our simulations show that the
exponential shape of the mass--velocity distribution in the outflow
is a rather general result.  We note that distributions are somewhat
irregular when considering only the ambient material ($j_{mix}$=0).
As soon as some degree of mixing is allowed between the jet and the
ambient gas, an exponential fitting provides a very satisfying
solution to the mass--velocity distribution at all positions. At
first sight, similar distributions are obtained for positions A1 to A3,
with values of $v_0\sim$ 5.1-6.6 km s$^{-1}$. These values are also
similar to those of the global outflow mass--velocity distribution,
as displayed in Fig.~\ref{f09}.

A higher value of $v_0$ of the order of 10 km s$^{-1}$ is found
at position A4, close to the apex of the outflow cavity. Hence, it
appears that if we leave aside the head of the outflow, the
mass--velocity distributions display only modest variations across
the outflow cavity, and do not bear signatures of the ejection
process (intermittency, precession).

A closer look at the distributions shows that the mass--velocity
distributions are better described by two components towards
positions A1 to A3, with a change of slope (index) near $v= 25\kms$.
In the high-velocity range, a shallower distribution is observed,
which is related to the jet-entrained material.  In order to explore the sensitivity of the fitting parameters to the geometry
and the age of the outflow, we extracted  the mass--velocity
relationships at positions A1-A4 at three different times in  the
simulation of model DR\_P, namely 400, 1200, and $2000\yr$, and for three
values of the inclination angle with respect to the plane of the
sky: $i = $ 10, 30, and 60$^{\circ}$.  The fitting results of  the
mass--velocity relationships are summarized in Table~\ref{table_fit}.

\begin{table}
\small\addtolength{\tabcolsep}{-1pt}
\caption{Model DR\_P. Best exponential fitting  parameters to the
molecular mass--velocity distributions obtained  towards positions
A1--A4 in an aperture of 5000 au  at $t$= 400, 1200, and
$2000\yr$, for three values of inclination angle: 
$i$=10$^{\circ}$, 30$^{\circ}$, and 60$^{\circ}$.}
\centering
\renewcommand{\footnoterule}{}
\begin{tabular}{cc c c c c}
\hline\hline
$j_{mix}$ & Age     & Aperture & \multicolumn{3}{c}{$v_0$ ($\kms$)}   \\ [0.5ex] 
          & (years) &          & $i = 10^\circ$ & $i=30^\circ$ & $i=60^\circ$   \\ 
\hline
\hline
 0.0 &   400  &  A1   & 1.3 & 2.7  & 4.4   \\
     &        &  A2   & 1.3 & 2.6  & 4.4   \\
     &        &  A3   & 1.4 & 2.6  & 2.0   \\
     &        &  A4   & 1.1 & 1.3  & 1.9   \\
     &  1200  &  A1   & 3.8 & 4.7  & 5.2   \\
     &        &  A2   & 9.4 & 7.0  & 4.8   \\
     &        &  A3   & 10.0 & 10.2  & 3.9   \\
     &        &  A4   & 1.0 &  1.5 &  2.1  \\
     &  2000  &  A1   & 2.8 & 3.1  &  4.9  \\
     &        &  A2   & 2.7 & 4.2  &  4.9  \\
     &        &  A3   & 2.8 & 4.8  &  4.9   \\
     &        &  A4   & 3.2 & 5.9  &  1.8  \\
\hline
\multicolumn{3}{c}{$v_i \pm \sigma_{v_i}$ ($\kms$):} & 3.4 $\pm$ 3.1 & 4.2 $\pm$ 2.6 & 3.8 $\pm$ 1.4 \\
\hline
\hline
 0.3 &   400  &  A1   & 3.3 & 4.9  &  9.2  \\
     &        &  A2   & 3.2 & 4.8  &  10.6  \\
     &        &  A3   & 3.2 & 5.9  &  11.5  \\
     &        &  A4   & 4.3 & 6.0  &  13.8  \\
     &  1200  &  A1   & 2.2 & 5.1  &  7.2  \\
     &        &  A2   & 2.3 & 5.9  &  8.1  \\
     &        &  A3   & 2.6 & 5.2  &  11.6  \\
     &        &  A4   & 3.7 & 8.1  &  13.2  \\
     &  2000  &  A1   & 2.6 & 5.3  &  7.8  \\
     &        &  A2   & 2.4 & 5.9  &  7.9  \\
     &        &  A3   & 2.7 & 4.4  &  8.4  \\
     &        &  A4   & 2.5 & 5.5  &  15.0  \\
\hline
\multicolumn{3}{c}{$v_i \pm \sigma_{v_i}$ ($\kms$):} & 2.9 $\pm$ 0.6 & 5.6 $\pm$ 1.0 & 10.4 $\pm$ 2.6 \\
\hline
\hline
 0.6 &   400  &  A1   & 3.2 & 6.5 &  9.6  \\
     &        &  A2   & 3.2 & 6.5 &  11.3   \\
     &        &  A3   & 3.1 & 7.3 &  13.0   \\
     &        &  A4   & 4.0 & 9.0 &  15.5   \\
     &  1200  &  A1   & 2.8 & 5.6 &  7.2   \\
     &        &  A2   & 2.9 & 6.3 &  8.2   \\
     &        &  A3   & 2.9 & 6.6 &  11.1   \\
     &        &  A4   & 3.6 & 10.6 & 14.2    \\
     &  2000  &  A1   & 2.8 & 5.5 &  7.8   \\
     &        &  A2   & 3.0 & 6.1 &  8.0   \\
     &        &  A3   & 2.9 & 5.6 &  8.6   \\
     &        &  A4   & 2.7 & 6.6 &  15.2   \\
\hline
\multicolumn{3}{c}{$v_i \pm \sigma_{v_i}$ ($\kms$):}  & 3.1 $\pm$ 0.4 & 6.9 $\pm$ 1.5 & 10.8 $\pm$ 3.0 \\
\hline
\hline
 1.0 &   400  &  A1   & 5.9 & 6.6 &  9.6   \\
     &        &  A2   & 6.2 & 6.5 &  11.3   \\
     &        &  A3   & 7.0 & 7.3 &  13.0   \\
     &        &  A4   & 6.5 & 9.0 &  15.5   \\
     &  1200  &  A1   & 4.8 & 5.6 &  7.2   \\
     &        &  A2   & 4.1 & 6.3 &  8.2 \\
     &        &  A3   & 4.5 & 6.6 &  11.1   \\
     &        &  A4   & 4.4 & 10.6 & 14.2    \\
     &  2000  &  A1   & 4.5 & 5.5 &  7.8   \\
     &        &  A2   & 4.7 & 6.1 &  8.0   \\
     &        &  A3   & 3.9 & 5.6 &  8.6   \\
     &        &  A4   & 3.4 & 6.7 &  15.2   \\
\hline
\multicolumn{3}{c}{$v_i \pm \sigma_{v_i}$ ($\kms$):}  & 5.0 $\pm$ 1.1 & 6.9 $\pm$ 1.5 & 10.8 $\pm$ 3.0 \\
\hline
\hline
\end{tabular}
\label{table_fit}
\end{table}

\section{Discussion}\label{discussion}

\subsection{Molecular outflows}

\cite{Lefloch2012} modeled  the observed intensity--velocity
distribution $I_{\rm CO}$ of the five outflow sources previously
observed by \cite{bachiller99}: Mon R2, L1551,  NGC2071, Orion A, and
L1448.  These latter authors showed that the observed  $I_{\rm CO}$ could
be  well fitted by an exponential function with
values of $v_0$ between 1.6 (Mon R2) and 12.5 (L1448) $\kms$.

\begin{figure*}
\centering
\includegraphics[width=1.\columnwidth]{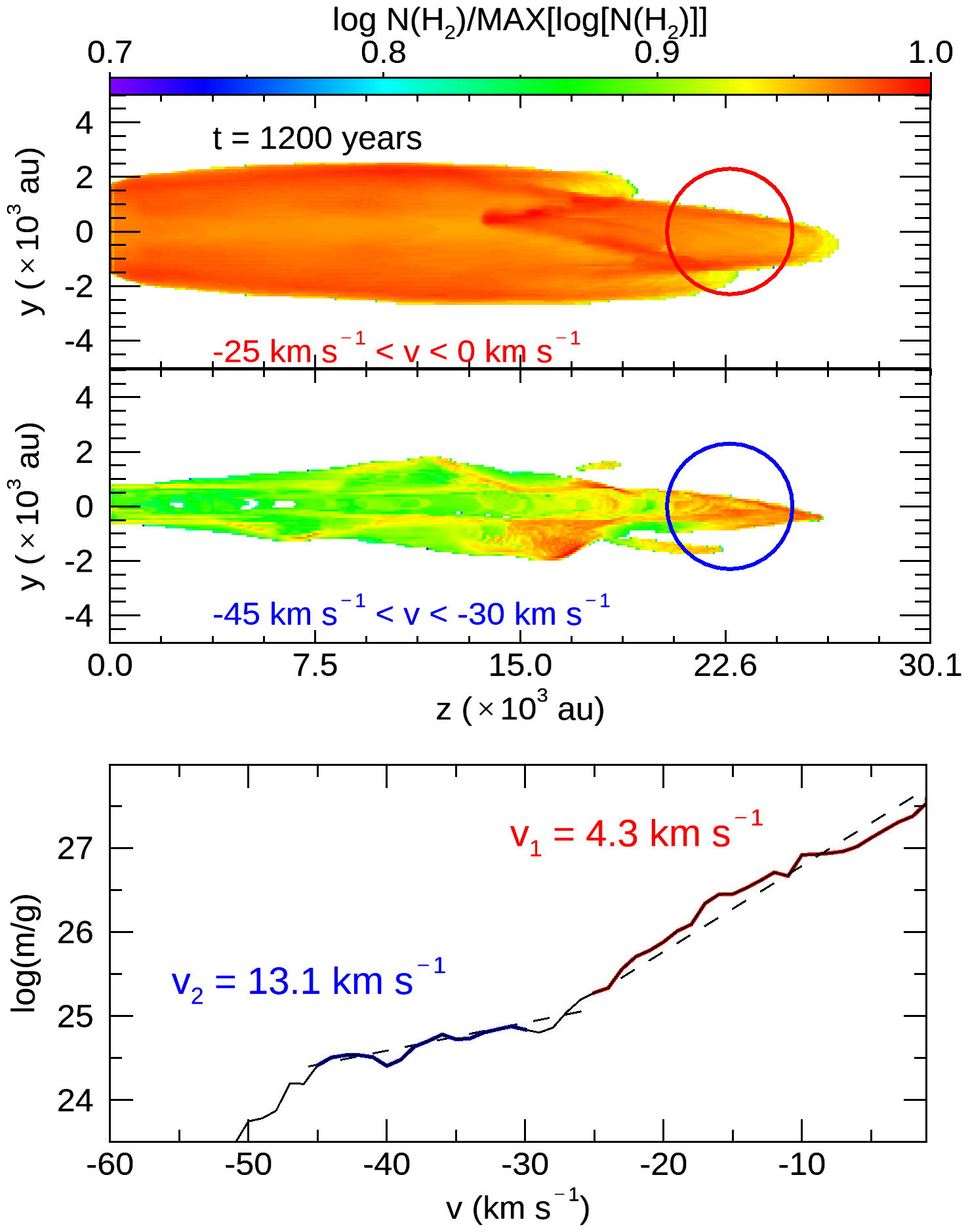}
\includegraphics[width=\columnwidth]{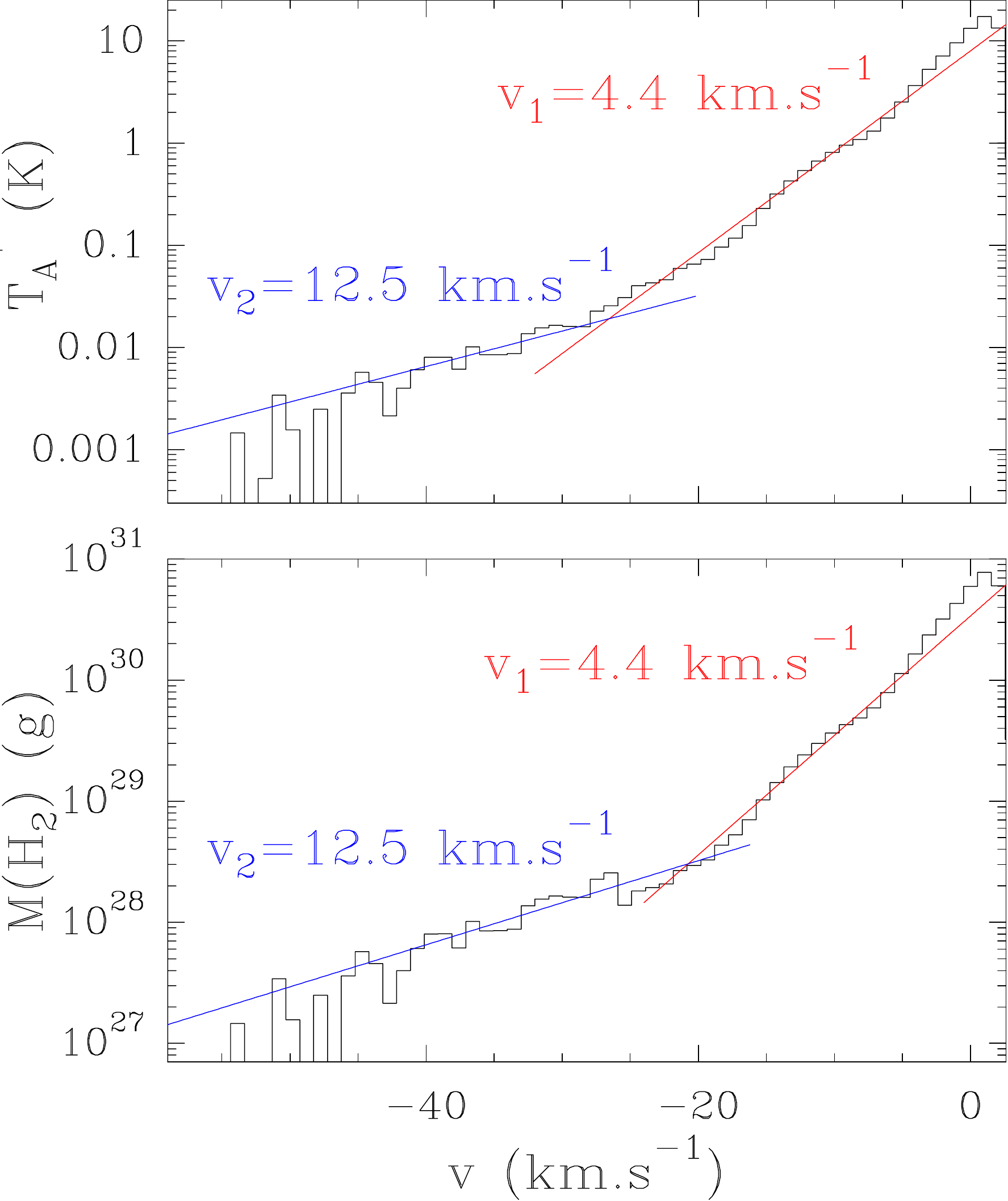}
\caption{{\em(Left)}~Model DR\_P at $t$=$1200\yr$ under an inclination
of $10^{\circ}$ (towards the observer).  {\em (top)}~Map of H$_2$
column density  for $j_{mix}= 0.9$, in the velocity intervals
$-[1-25\kms]$ (top) and $-[30-45\kms]$ (bottom). The 5000 au aperture
used to extract the mass--velocity distribution is drawn by a circle.
{\em (bottom)}~Molecular mass--velocity distribution extracted at
position A4.  We have superposed the best exponential fits $m(v)\propto
\exp(-v/v_i)$ to the components associated with velocity intervals
$-[1-25\kms]$ (red) and $-[30-45\kms]$ (blue). The exponent value
$v_i$ is given for both velocity intervals.  {\em (Right)}~ASAI
observations of L1157-B1.  {\em (top) }Intensity--velocity distribution
obtained in the CO $J$=2--1 towards shock position B1 in an aperture
of about 4000 au (11\arcsec at the distance to the source).  The
line profile is fitted by a linear combination of two exponential
functions $g_1\propto \exp(-v/12.5)$ (blue), $g_2\propto \exp(-v/4.4)$
(red) \citep[from][]{Lefloch2012}. {\em (bottom)}~Associated
mass--velocity distribution, adopting a standard CO-to-\htwo\
abundance ratio and the excitation conditions derived by \cite{Lefloch2012}
for the velocity components $g_1$ and $g_2$.}\label{f11}
\end{figure*}

A quick inspection of the grid of models presented in Table~\ref{table_fit}
shows that these values fall well within the range of values predicted
in our simulations, depending on the outflow age, the inclination
angle, and the degree of  jet--ambient gas mixing ratio.  A value as
low as 1.6 for Mon R2 is indeed easily accounted for if the outflow
propagates close to the plane of the sky, as proposed by DC03.  For
an inclination angle of $10^{\circ}$, and a time of 1200-$2000\yr$,
our modeling predicts low values $v_0\sim 1.5-2.0$, which are
easily obtained where there is a lack of entrainment in the outflowing
gas ($j_{mix}$=0.0). We note that these values are not very sensitive
to the actual age of the outflow (time of the simulation).

The young sources Orion A and L1448, with an estimated age of
$1000\yr$, are best fitted with $v_0$ values of  $6.6$ and $12.5
\kms$ \citep[according to][]{Lefloch2012}. These values of $v_0$
are easily accounted for in the simulations  at early ages
(400-$1000\yr$) with an inclination angle of 60$^{\circ}$.  Solutions
with a lower inclination angle and a different degree of jet--ambient
gas mixing ratio are possible. Detailed modeling of the sources
is necessary to disentangle the impact of the different parameters.
Interestingly, the spectra for L1448 presented in Fig.~4 in
\cite{Lefloch2012} present a bump at $v\sim 60\kms$, which we
interpret as  the signature of the driving jet.

From comparison with the outflow sample of \cite{bachiller99}, a
scenario emerges in which more-evolved sources (like Mon R2) are
better described by the swept-up gas (i.e, no entrainment; $j_{mix}=0$),
while younger sources unveil the ambient medium entrained by the
jet, making the profiles shallower and consistent with $v_0 \sim
10 \kms$.

\subsection{Mass--velocity relationships in the L1157 outflow}

In this section, we apply our numerical models to L1157 in order
to better understand the origin of the CO intensity--velocity
distributions reported by \cite{Lefloch2012} in the southern outflow
lobe of L1157. We note that our goal  is not to provide a detailed
modeling of the L1157 southern outflow lobe.  For this reason, we
focus on the signatures associated with the B1 outflow cavity.

As mentioned in Section \ref{simu}, the simulation parameters of
model DR\_P were chosen to describe the behavior of a "typical"
precessing jet.  For this reason, and taking into account the
simplicity of the underlying hypothesis of our model, we did not
attempt to fine-tune the simulation parameters in order to obtain
the "best-fitting" model.

\subsubsection{Multiple components}

As mentioned in Section \ref{newsec2}, several authors have investigated the
details of the  CO emission from the southern lobe of the L1157 outflow.
As first shown by \cite{gueth1996}, the precessing protostellar jet
has shaped the southern  lobe into two shells (outflow cavities)
whose apexes are associated with the molecular shock  positions B1
and B2.  Detailed modeling of the CO gas kinematics by \cite{podio2016}
showed that the jet precesses on a cone  inclined by $73^{\circ}$
to the line of sight, with an opening angle of $8^{\circ}$ on a
period of $1640\yr$.  The modeling of the authors indicates that
 an angle of $\approx 10^{\circ}$ exists between the jet and the line of sight
at the location of B1.

The top-right panel of Fig.~\ref{f11} shows the CO
intensity--velocity distribution as observed in the $J$=2--1 line
with IRAM-30m \citep{Lefloch2012}.  From a multi-transition
analysis of the CO line profiles, these latter authors found  the following
results:

\begin{itemize} 
\item The CO line profiles profiles are the sum of up to three
components of specific excitation and velocity range, dubbed $g_1$,
$g_2$, $g_3$, all of which can be modeled by an exponential law
with a specific exponent $v_0$: 12.5, 4.4,  2.5, respectively.
\item The component of lowest excitation ($T_{ex}$=$23\K$) and
narrowest velocity range, ($-5;0\kms$), dubbed $g_3$, is detected
over the whole southern lobe and is the only component detected
towards the southernmost, older cavity associated with the B2
shock.
\item The component of highest excitation ($T_{ex}$=$210\K$)  and
highest velocity range, $[-40;-20]\kms$, dubbed $g_1$, is detected
close to the apex of the younger cavity associated with the B1 shock.
\item The component of intermediate excitation ($T_{ex}$=$64\K$)
and velocity range, $-20 < v < -5\kms$, dubbed $g_2$ is detected
over the whole outflow cavity associated  with B1.
\end{itemize}

\subsubsection{Observational derivation of m(v)}

The excitation conditions of the CO gas in L1157 make it especially
easy to obtain the mass--velocity distribution from the CO
intensity--velocity distribution. This is because the  CO $J$=2--1
excitation conditions of each component  $g_1$, $g_2$, $g_3$ are
independent of the velocity and the line emission is optically thin,
but at velocities very close (a few $\kms$) to ambient.  Hence,
the simple relation that exists at local thermodynamic equilibrium between N(CO) and the CO
line flux \citep[see e.g.,][]{bachiller90} can be applied to the
whole velocity range of emission of each component.  In practice,
each component dominates over a specific velocity interval of the
intensity--velocity distribution (see top right panel of Fig.~11).
The mass--velocity distribution is therefore immediately obtained
when considering the excitation conditions and the size of the main
emitting gas component as a function of velocity. The total
mass--velocity relationship is rigorously obtained by multiplying
the relationship N(CO)(v) ---which is derived from $T_b$(CO)(v)--- by the CO
emission area at each velocity interval. Despite the uncertainties
in the overlap region between components (e.g., near $v=-25\kms$),
 the spectral slope of each component is found to be
preserved in the derivation procedure from the CO intensity to the
mass--velocity distribution.  This is illustrated in the bottom-right
panel of Fig.~\ref{f11} in which we report the mass--velocity
distribution towards L1157-B1. We note that we assume a standard
abundance ratio  CO/[$\htwo$]= $10^{-4}$.

\subsubsection{Spatial distribution}

Figure~\ref{f11} presents the distributions of the molecular
material in the velocity intervals $[-25;-1]\kms$ (top) and
$[-45;-30]\kms$, respectively.  We also computed the molecular
mass--velocity distribution measured in an aperture of 5000 au close
to the apex of the cavity (position A4).  This is comparable to the
beamwidth (HPBW) of the IRAM 30m telescope main beam at the frequency
of the CO $J$=2--1 line.

Our simulations (see Fig. \ref{f10}) show that after a few hundred years the intensity--velocity distribution is approximately uniform over
the outflow cavity, except at the apex where the jet contribution
is strongest. This is consistent with the detection of rather uniform
signatures $g_2$ and $g_3$ over the B1 and B2 lobes, respectively.
Our model DR\_P is consistent with the interpretation that   $g_2$
and $g_3$ are associated with different ejection events responsible
for the formation of B1 and B2 cavities, respectively. The lower
excitation conditions of $g_3$ are  consistent with an  older event.
The difference of spectral slope ($v_0$) between the B1 and B2
cavities could be explained {\em a priori} by  a higher jet inclination
to the line of sight in the direction of B2. However, this contradicts
the kinematic modeling of the jet precession by \cite{podio2016},
which predicts an inclination angle to the line of sight of about
$25^{\circ}$ at shock position B2, lower than towards B1, and therefore
favors a higher jet radial velocity than that measured towards B1.
The sensitivity of the millimeter CO line spectra  available in the
literature \citep[see e.g.,][]{Lefloch2012} is not high enough to allow conclusions to be made about the presence of the jet towards B2. On the other
hand, our numerical simulations show that the low-velocity ($v <
5\kms$) emission actually arises from entrained ambient material
in the outflow cavity walls, and has very little dependence  on the driving jet.
The jet that once created the B2 shock about $2500\yr$ ago\footnote{We adopted the revised distance of $372\pc$ to L1157 \citep{zucker2019}.}
\citep{podio2016} is now impacting the B1 cavity at the B1 and B0
positions as a result of its precession.  Hence, the emission from
B2 arises mainly from previously entrained, ambient material, which
is now being slowed down. Inspection of Table~\ref{table_fit} shows
that low values of $v_0$ are also obtained in the swept-up ambient
molecular gas ($j_{mix}= 0.0$).

\subsubsection{Jet signature}

As can be seen in Fig.~\ref{f11}, the mass--velocity distribution
extracted towards position A4, close to the apex of the cavity in
the simulation, shows the presence of two distinct components
associated with the velocity intervals [$-1;-25\kms$] and
[$-30;-45\kms$], respectively. This situation is reminiscent of the
CO (and the mass) intensity--velocity distribution observed towards
B1.  Both components can be fitted by an exponential function of
exponent $v_0\simeq 4.3$ and $v_0\simeq 13$, respectively.  These
values are in good agreement with those determined for
components $g_2$ and $g_1$  towards the B1 position.  We note that
according to our modeling (see Table~\ref{table_fit}), these values
are mainly sensitive to the jet inclination angle to the line of
sight. We note that they weakly depend on the actual value of the
jet--ambient material mixing ratio, but the best agreement was obtained
for $j_{mix}$= 0.9.

In our simulation, the distribution of the high-velocity
material between -45 and $-30\kms$, as displayed in Fig.~\ref{f11},
is not restricted to a few spots of shocked gas, such as for example the
jet impact shock region at the apex of the outflow cavity.  Instead,
it turns out that the high-velocity material ($ \vert v \vert  >
30\kms$) is tracing an elongated, collimated structure surrounding
the jet throughout the whole outflow cavity. This elongated structure
is surrounded by the lower velocity material ($0 < \vert v \vert <
25\kms$) of the outflow. In other words, the high-velocity material
does not trace only the jet shock impact region (the Mach disk).
In our simulation, the amount of molecular material at the jet head
strongly decreases as a result of molecular dissociation. Instead,
the high-velocity component arises from material entrained
along the jet.

\subsubsection{Observational predictions for the high-velocity gas}

Observational evidence of the high-velocity component has been
reported in the millimeter rotational transitions of a few
molecular species, such as for example CO \citep{Lefloch2012}, HCO$^{+}$
\citep{podio2014}, and SiO \citep{tafalla2010,spezzano2020}.
Unfortunately, the interferometric observations of L1157 available
in the literature focus mainly on the gas propagating at
velocities close to ambient, which is associated with the bow and the outflow
cavity walls.  Therefore, the evidence for the high-velocity jet is
still very scarce and unambiguous detection of the molecular material
entrained along the jet is still missing. It is worth noting that
Plateau de Bure observations of the SiO $J$=2--1 line by \cite{gueth1998}
revealed an elongated , "jet and filamentary-like" feature for gas
emitting at $v < -10\kms$. Single-dish observations of the SiO
$J$=2--1 line indicate that this feature displays the expected
intensity--velocity distribution, as shown by \cite{Lefloch2012}.
We speculate that this feature might well be the signature of the
molecular jet and not the jet impact shock region itself.

This conclusion could be easily verified (or disproved) by high-angular
resolution observations of the CO or SiO  millimeter line emission
with  the IRAM interferometer NOEMA. According to our modeling,
the high-velocity emission should reveal a collimated, filamentary-like
structure. In the opposite case, that is, if the gas is accelerated in
the jet impact shock region, the high-velocity emission should trace
the compact region associated with the Mach disk.

\section{Conclusions}\label{conc}

Using the hydrodynamical code Yguazú-a, we performed 3D numerical
simulations to revisit in a detailed manner the mass--velocity
relationship in jet-driven molecular outflows.  Great attention was
paid to benchmark Yguazu-a against the hydrodynamical codes used
by previous authors in the field \citep{DR99, downes}. To do so,
we  modeled the propagation of an intermittent jet adopting  the
same parameters as those of \cite{DR99} and \cite{downes}.  We find excellent
quantitative agreement between our simulations and those of  these latter
authors.

Detailed comparison between our simulations and those of  \cite{downes} leads us to
conclude that these latter authors took into account the jet material
contribution in the obtention of the mass--velocity distribution
presented in their work.   We find that the presence of a bump in
the high-velocity range ($v \sim 100 \kms$) is remarkable evidence
of the presence of the jet and that all the previous works
considered, to a greater or lesser extent, the presence of the jet in their
mass--velocity profile computations.

Overall, our simulations show that the mass--velocity
distribution of the outflowing material can be successfully fitted
by one exponential law $m(v)\propto \exp(-v/v_0)$. We systematically investigated
the signature of the mass--velocity distribution
as a function of time, depending on the jet inclination to the line
of sight and the degree of mixing between the jet and the ambient
material. We find that it may be necessary to introduce a second
component to fit the mass--velocity distribution in the high-velocity
range. The spectral signature in the  low-velocity range is dominated
by the contribution of material in the outflow cavity walls and is
rather insensitive to the  actual value of the jet--ambient gas
mixing ratio. The synthetic mass--velocity distributions from  our
simulations are good agreement with distributions derived from observations and we are
able to reproduce the observational data when taking age and source geometry  into account.

We verified that the profile of the mass--velocity distribution
computed over a local area inside the outflow can still be well fitted by an exponential function. The profiles and the $v_0$ values
are very similar over the outflow, but  the distribution appears much shallower at the apex of the outflow
cavity as a result
of the leading jet contribution.

We performed a simple modeling of the L1157 southern outflow
cavity  by simulating of a precessing jet with parameters similar
to those reported by \cite{podio2016}. We were able to reproduce
the main features of the CO intensity--velocity distributions  observed
in the southern outflow lobe of L1157 to a satisfactory degree. Our simulations suggest that
the three components identified by \cite{Lefloch2012} are all related
to the entrained gas and not the jet impact shock region, that is, the
Mach disk itself.  High-angular-resolution observations with the
NOEMA interferometer could easily test this conclusion.

\begin{acknowledgements}
We would like to thank to an anonymous referee, whose suggestions
has contributed to improve the presentation of the paper. A.H.
Cerqueira would like to thanks the PROPP--UESC for partial funding
(under project No. 073.6766.2019.0010667-91) as well as the Brazilian
Agency CAPES.  B. Lefloch, P.R.  Rivera-Ortiz and C. Ceccarelli
acknowledge funding from the European Research Council (ERC) under
the European Union's Horizon 2020 research and innovation programme,
for the Project “The Dawn of Organic Chemistry” (DOC), grant agreement
No 741002.
\end{acknowledgements}


\end{document}